\documentclass[letterpaper, 10 pt, conference]{ieeeconf}
\IEEEoverridecommandlockouts                              
\usepackage{amsmath,amssymb,graphicx,bm}
\usepackage{cite}
\usepackage{float}
\usepackage{subcaption}
\usepackage{graphicx}
\makeatletter
\renewcommand*\env@matrix[1][\arraystretch]{%
  \edef\arraystretch{#1}%
  \hskip -\arraycolsep
  \let\@ifnextchar\new@ifnextchar
  \array{*\c@MaxMatrixCols c}}
\makeatother
\title{\LARGE \bf
Evaluation of A Semi-Autonomous Lane Departure Correction System Using Naturalistic Driving Data
}
\author{Ding Zhao$^{1}$, Wenshuo Wang$^{2}$, David J. LeBlanc$^{1}$ 
\thanks{*This work was supported by Toyota Class Action Settlement Safety Research and Education Program}
\thanks{$^{1}$D. Zhao (corresponding author: {\tt\small zhaoding@umich.edu}) and D. J. LeBlanc are with the University of Michigan Transportation Research Institute, Ann Arbor, MI 48109}%
 \thanks{$^{2}$W. Wang is with the Dept. of Mechanical Engineering, Beijing Institute of Technology, Beijing, China 100081 and the Dept. of Mechanical Engineering, University of California at Berkeley, Berkeley, CA, 94720 USA.}
}

\begin{document}

\maketitle
\thispagestyle{empty}
\pagestyle{empty}

\begin{abstract}
Evaluating the effectiveness and benefits of driver assistance systems is essential for improving the system performance. In this paper, we  propose an efficient evaluation method for a semi-autonomous lane departure correction system. To achieve this, we apply a bounded Gaussian mixture model to describe drivers' stochastic lane departure behavior learned from naturalistic driving data, which can regenerate departure behaviors to evaluate the lane departure correction system. In the stochastic lane departure model, we  conduct a dimension reduction to reduce the computation cost. Finally, to show the advantages of our proposed evaluation approach, we compare steering systems with and without lane departure assistance based on the stochastic lane departure model. The simulation results show that the proposed method can effectively evaluate the  lane departure correction system.
\end{abstract}
\begin{keywords}
Performance evaluation, lane departure correction system, stochastic driver model, bounded Gaussian mixture model
\end{keywords}

\section{Introduction}
In the United States, single-vehicle road departures accounted for approximately 37.4 \% of all fatal vehicle crashes. Many studies have been conducted on the lane departure warning (LDW) system, lane departure assistance (LDA) system, and lane centering assistant (LCA) system. These systems have the potential to address a large proportion of serious injury and fatal crashes and have been studied by many authors. For example, Minoiu Enache \cite{enache2009driver} \textit{et al}.,  designed a steering assistant controller for lane departure behaviors. Reagan and McCartt \cite{reagan2016observed} investigated on the frequency in which the LDW system was activated. A limited number of studies, however, have been conducted on how to evaluate the effectiveness and benefits of these systems in the real world. Several naturalistic field operational tests have been conducted in the U.S. However, the cost of these types of evaluation is so high that they are useful only for evaluating the final product, not at the point of development. Mathematical models capable of reproducing drivers' lane departure behaviors can be applied to testing and evaluating the effectiveness and benefits of these systems, thus cutting costs and shortening the development cycle. Such models may also allow deeper insight into the physiological and cognitive behaviors of human drivers so that present or future driver-automation interfaces can be optimized. 

\begin{figure}[t]
	\centering
	\includegraphics[width=\linewidth]{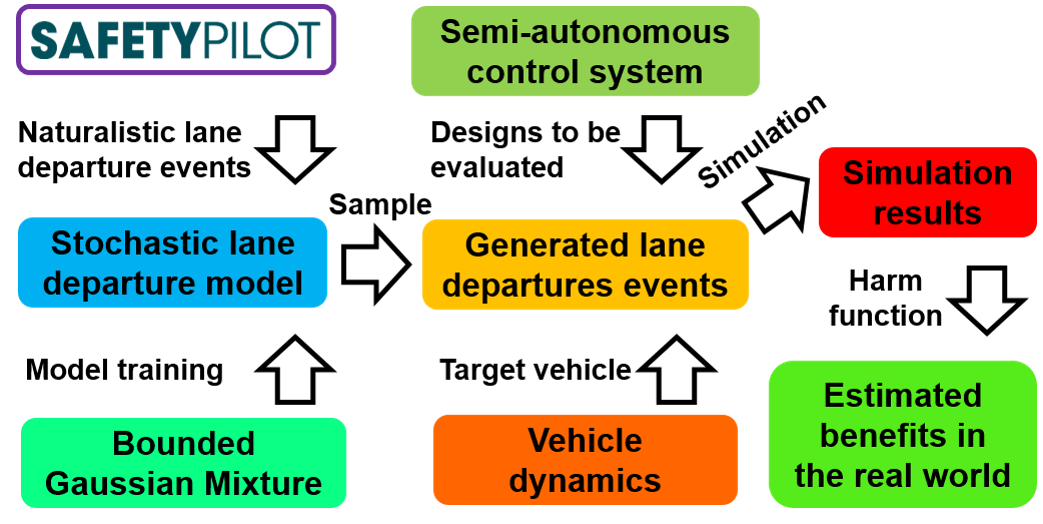}
	\caption{Evaluation procedure based on a stochastic lane departure model using naturalistic driving data.}
	\label{fig:procedure}
\end{figure}

Recently, simulation-based evaluation methods have become popular \cite{Zhao2016AcceleratedTechniques, Zhao2015AcceleratedData, Zhao2016AcceleratedManeuvers,harding2014vehicle,wanghuman,Huang2016UsingScenario} in evaluating the safety of automated vehicles and the performance of advanced driver assistance systems. In this paper, we propose a simulation-based method to evaluate the lane departure correction (LDC) system. The evaluation procedure is shown in Fig. \ref{fig:procedure}. First, a stochastic lane departure model is built based on a bound Gaussian mixture  (BGM) model using a very large quantity of naturalistic driving data. In the trained model, we use a dimension reduction method and apply 8 variables representing the lane departure behavior. Thus, based on the stochastic lane departure model, we can extract and regenerate the lane departure event as similar to what it might be in the real word. The lane departure events generated from the stochastic model can be used to test and evaluate the LDC system. When the vehicle generated from the stochastic model drifts  cross the lane marker, the LDC system will automatic control the power steering, thus bringing the vehicle back into the lane. In this way, we can estimate different LDC systems.

\section{Modeling Lane Departure Behaviors}
\subsection{Naturalistic Lane Departure Events}
The naturalistic driving data used in this research are extracted from the Safety Pilot Model Deployment (SPMD) database \cite{bezzina2014safety}, which recorded the naturalistic driving of 2,842 equipped vehicles in Ann Arbor, Michigan, for over two years. As of April 2016, 34.9 million miles were logged, making the SPMD one of the largest public N-FOT databases ever. We used 98 sedans to run experiments and collect real on-road data. The vehicles were equipped with Data Acquisition System and Mobileye$^\circledR$ \cite{harding2014vehicle}. The Mobileye$^\circledR$ obtains the driving data such as relative range, relative speed, and lane tracking measures about lane delineation both from the painted boundary lines and the road edge, etc. The GPS obtains the global position (latitude and longitude) and the GPS time. The vehicle speed, acceleration, throttle opening, braking force, engine speed were obtained via the CAN bus. The error of range measurement to the front object was around 10 \% at 90 m and 5 \% at 45 m \cite{stein2003vision}.

To ensure consistency of the dataset, we applied the following criteria: (1) the duration of each event should be in the range of 0.5 s to 10 s and (2) the average velocity of each event should be larger than 5 m/s. In total, 529,096 lane departure events were identified from 118 drivers over the previous four years.
%
%
%

\subsection{Lane Departure Model}
\begin{figure}[t]
	\centering
	\includegraphics[width=\linewidth]{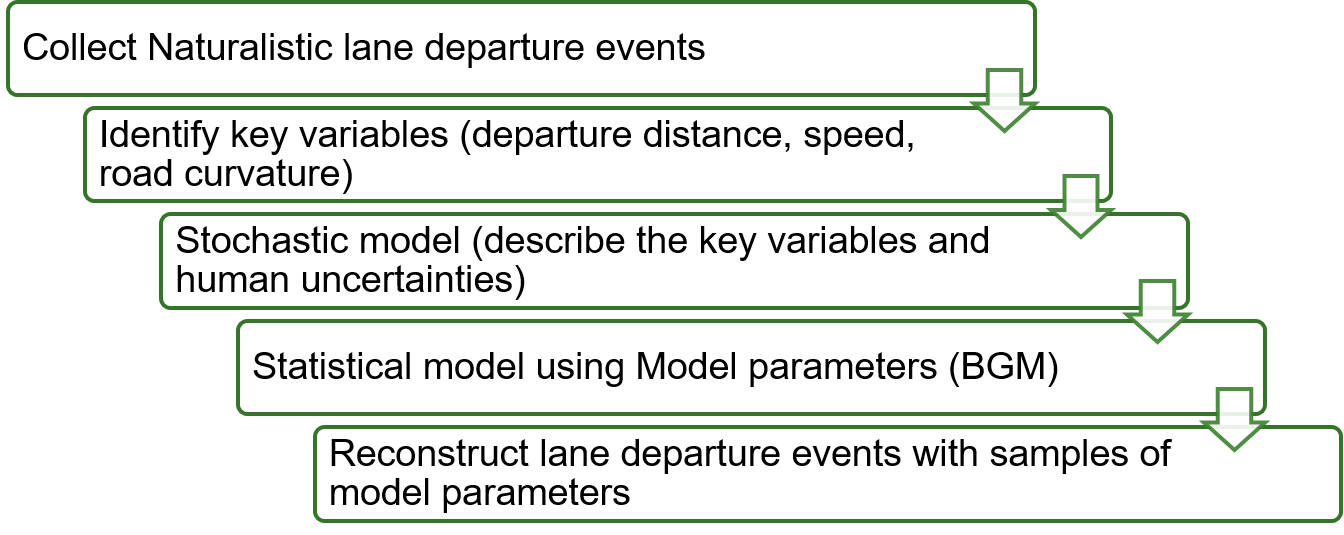}
	\caption{Procedure to build the lane departure model}
	\label{fig:modelProcedure}
\end{figure}
\begin{figure}[t]
	\centering
	\includegraphics[width=1\linewidth]{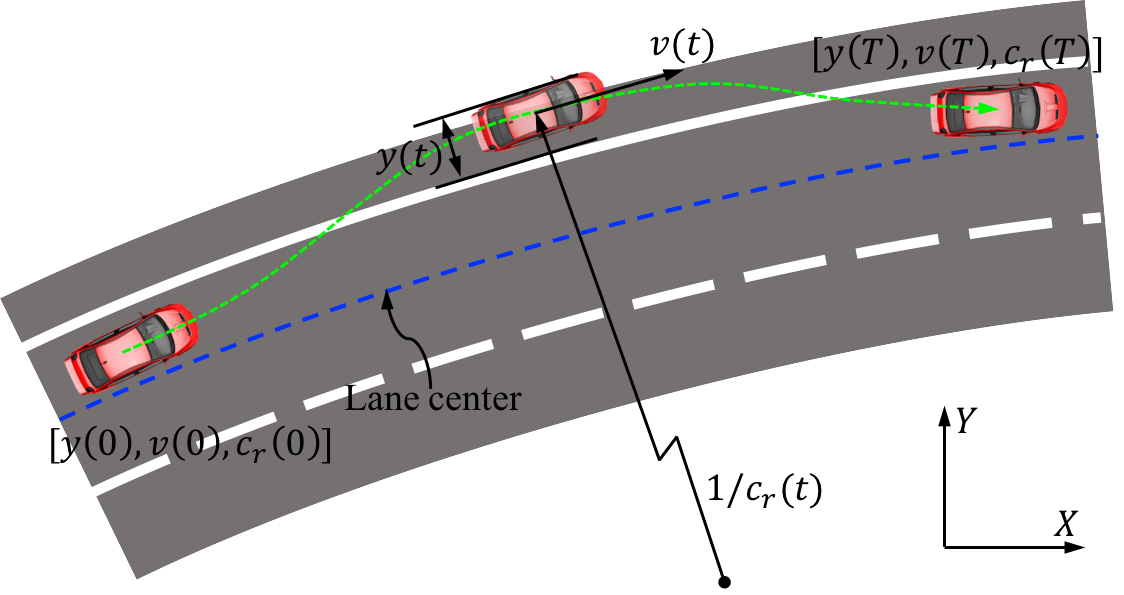}
	\caption{Illustration of lane departure model}
	\label{fig:LD_model}
\end{figure}
\begin{figure}[t]
      \centering
      \begin{subfigure}{0.5\textwidth}
      \includegraphics[width=\linewidth]{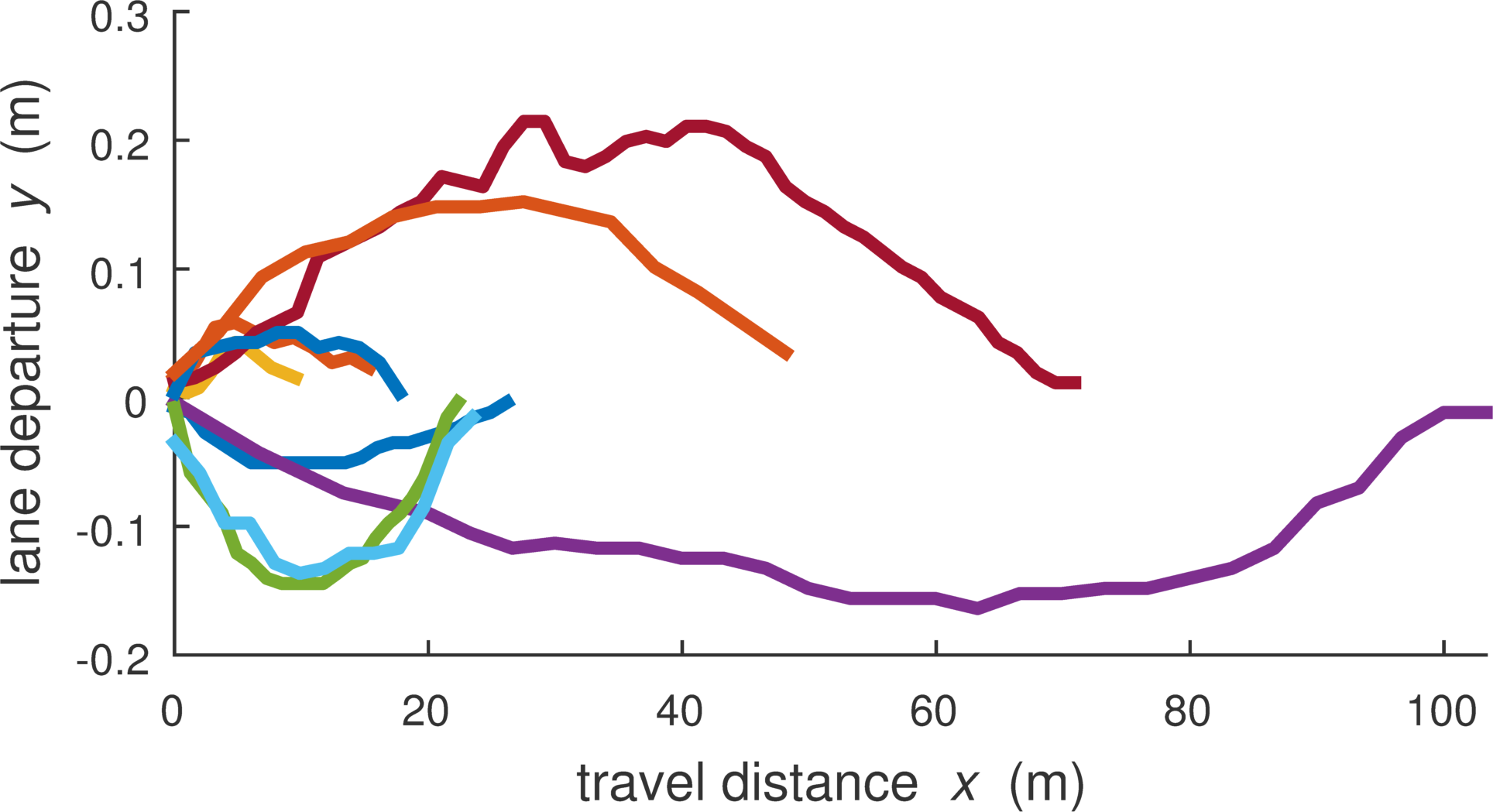}
      \caption{Lateral departure distance}
      \end{subfigure}
      \begin{subfigure}{0.5\textwidth}
      	\includegraphics[width=\linewidth]{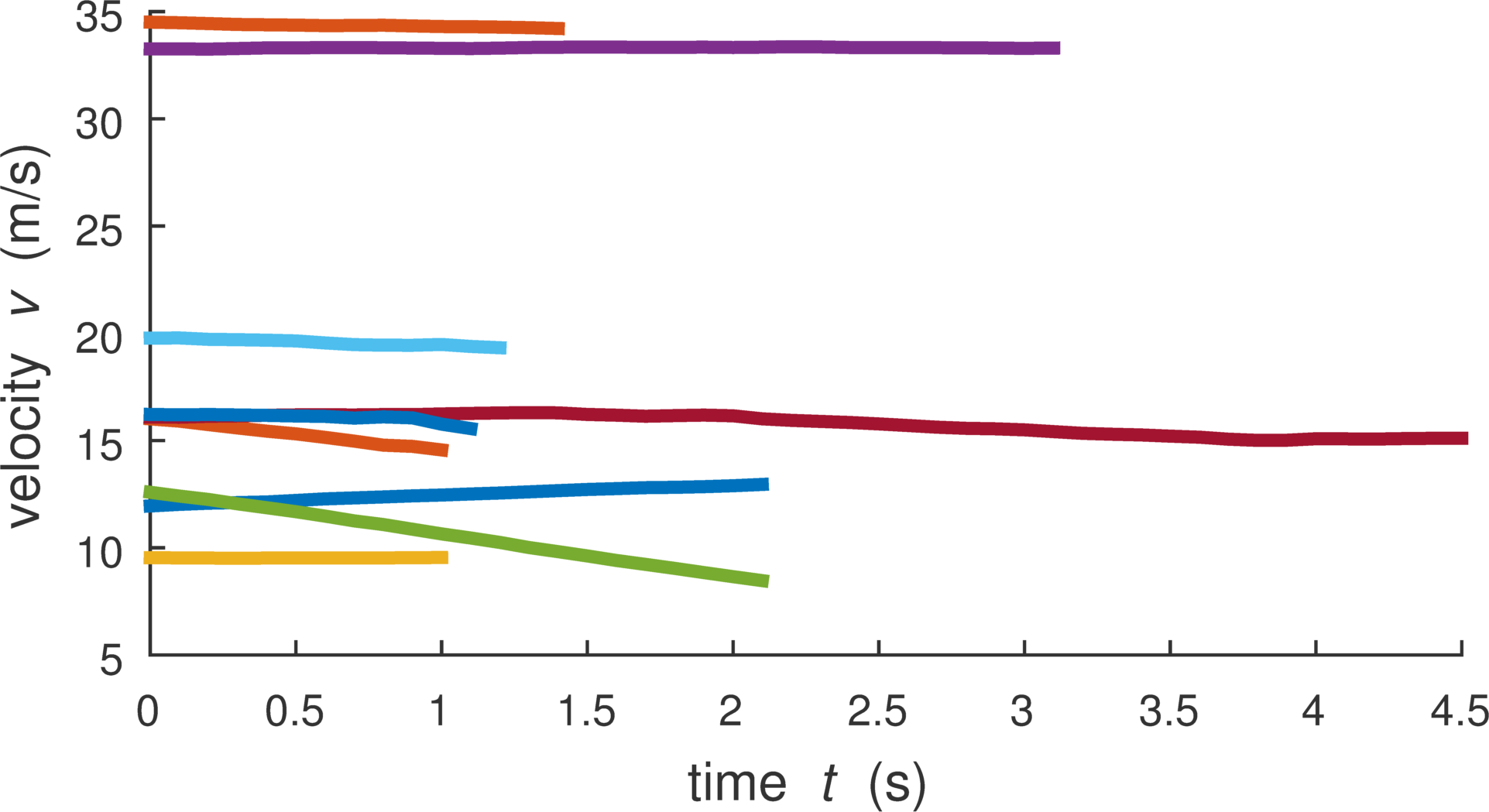}
      	\caption{Velocity}
      \end{subfigure}
      \begin{subfigure}{0.5\textwidth}
      	\includegraphics[width=\linewidth]{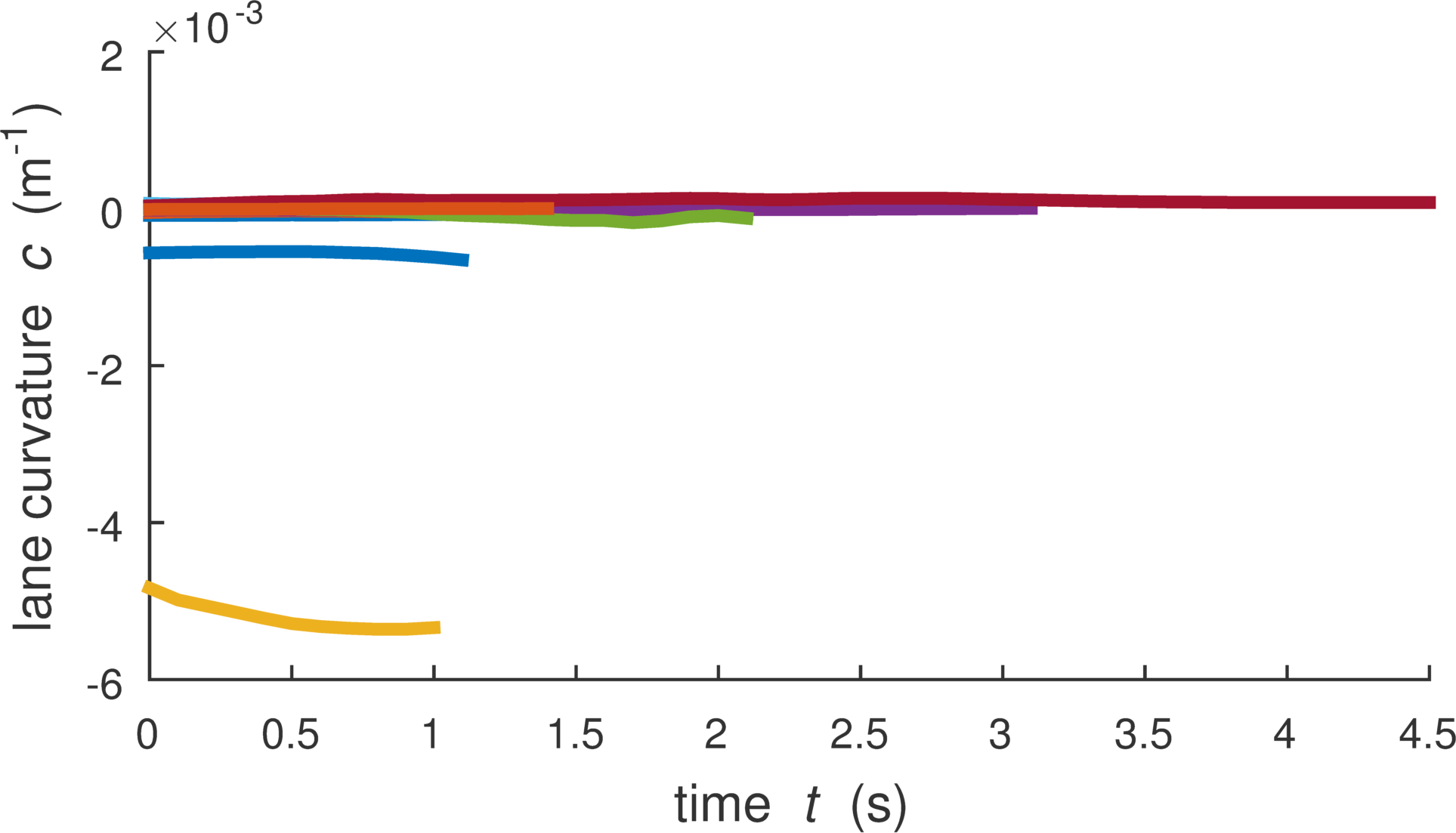}
      	\caption{Road curvature}
      \end{subfigure}
      \caption{Lateral offset examples during lane departure}
      \label{fig:variables}
\end{figure}

The model procedure is shown in Fig. \ref{fig:modelProcedure}. Three key variables are used to describe a lane departure event: (1) lane departure $y$, (2) vehicle speed $v$, and (3) lane curvature $c_l$ as shown in \ref{fig:LD_model}. Examples extracted from the SPMD database are shown in  Fig. \ref{fig:variables}. 

Our goal is to build a  model that can generate lane departure events statistically equivalent to the events collected from naturalistic driving data. If the departure duration $T$  is 5 s with sampling time $T_s$ being  0.1 s, we will get $3(T/T_s+1)=153$ data points to fully describe the three variables $y$  $y$, $v$, and $c_l$.  This dimension is normally too high to build a stochastic model. Therefore, the first task is to reduce the model dimension by extracting the key features of the variables while reserving model uncertainty. As illustrated in Fig. \ref{fig:variables}(a), $y$  can be approximated through a second order polynomial function of the longitudinal travel distance $x$ plus the error term.

\begin{equation}
y(t)=\tilde{y}(t)+\epsilon_y(t)
\end{equation}
\begin{equation}
\tilde{y}(t)=-\frac{4d_y}{d_x^2}\Big(x(t)-\frac{d_x}{2}\Big)^2+d_y
\end{equation}
where $d_x$ is the longitudinal travel distance during the departure. $d_y$ represents the lateral departure  calculated from the Least Square method.
Variance in human driving is captured by the standard deviation of  $\epsilon_y(l)$, which can be calculated from
\begin{equation}
\label{eq:sigma_y}
\sigma_y = \sqrt{\dfrac{1}{L-1}\sum_{l=1}^{L}|\epsilon_y(l)-\bar{\epsilon}_y|^2}
\end{equation}
where $ L $ is the number of samples in one event, $\epsilon_y(l)=y(l)-\tilde{y}(l)$ represents the error at the  time of the $l^{th}$ sample of the lane departure event. Without introducing confusion,  in the this paper we use $t$ to represent continuous time starting from 0 to T and use $l$ as the index of the discrete sample time, starting from 1 to L, and  $\bar{\epsilon}_y=\sum_{l=1}^{L}\epsilon_y(l)$.

\begin{figure}[t]
	\centering
	\begin{tabular}{cc}
		\includegraphics[width=0.47\linewidth]{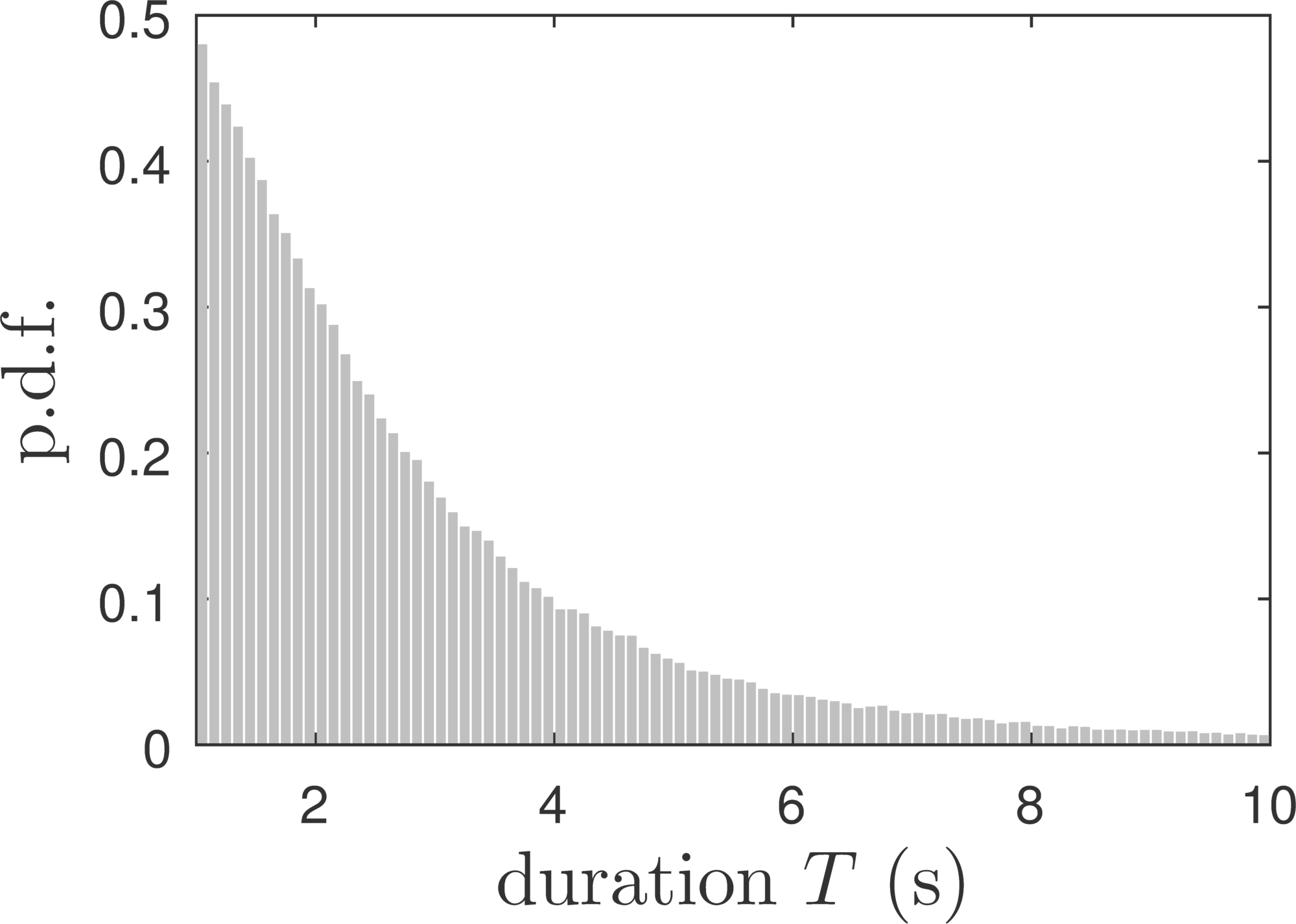} &
		\includegraphics[width=0.47\linewidth]{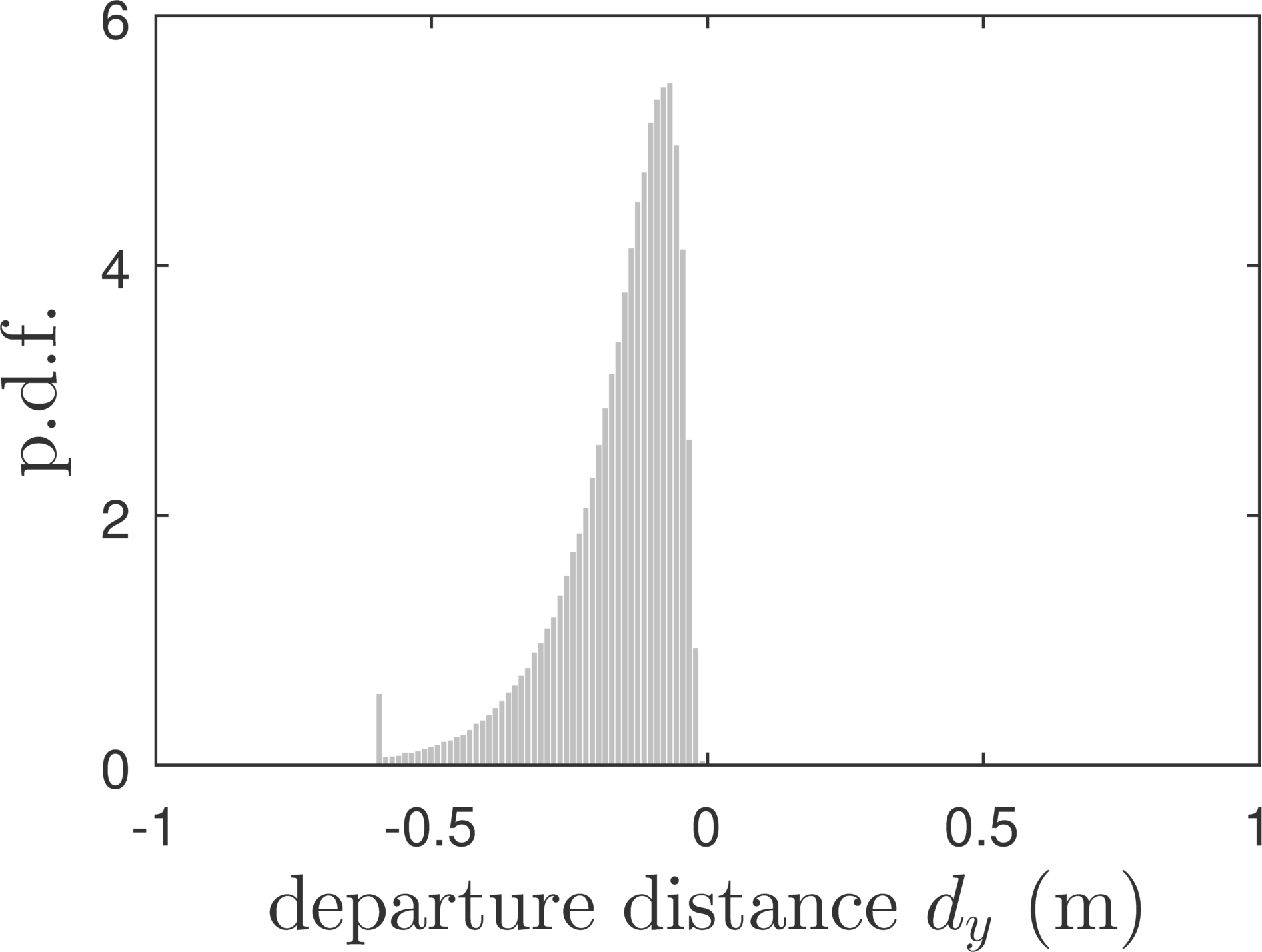}\\
		\begin{small} (a) \end{small} &\begin{small} (b) \end{small} \\
		\includegraphics[width=0.47\linewidth]{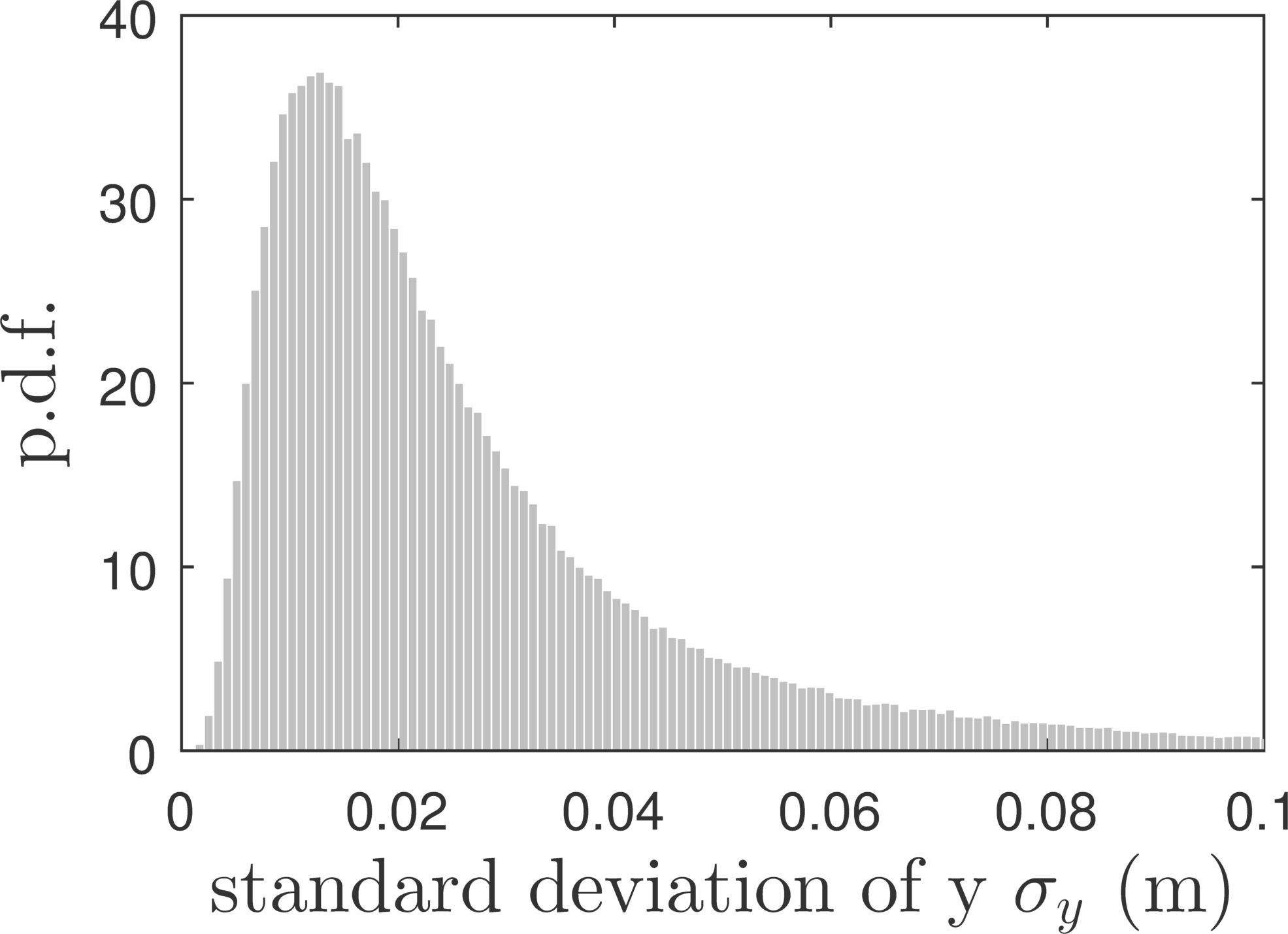}&
		\includegraphics[width=0.47\linewidth]{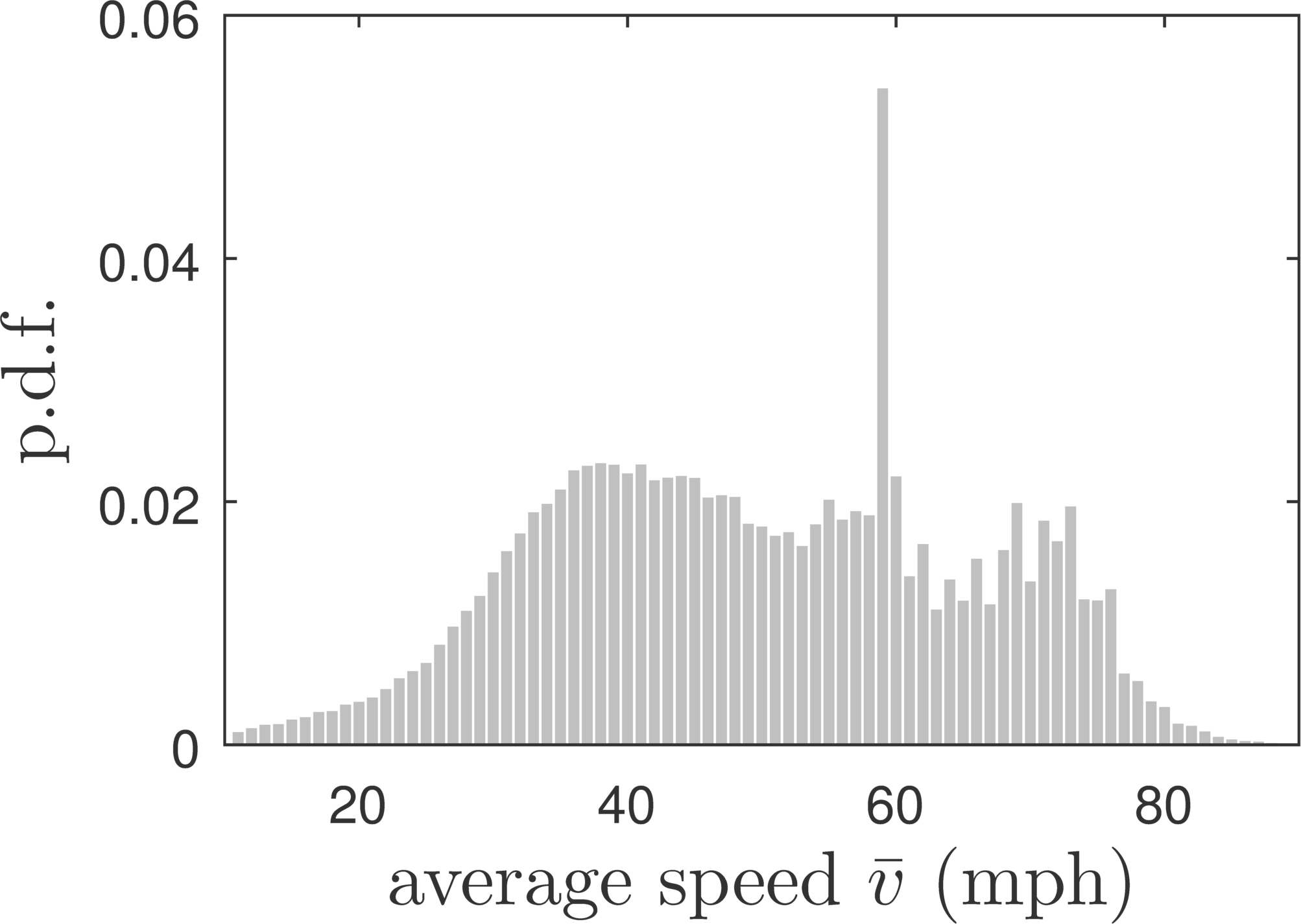}\\
		\begin{small} (c) \end{small} &\begin{small} (d) \end{small} \\
		\includegraphics[width=0.47\linewidth]{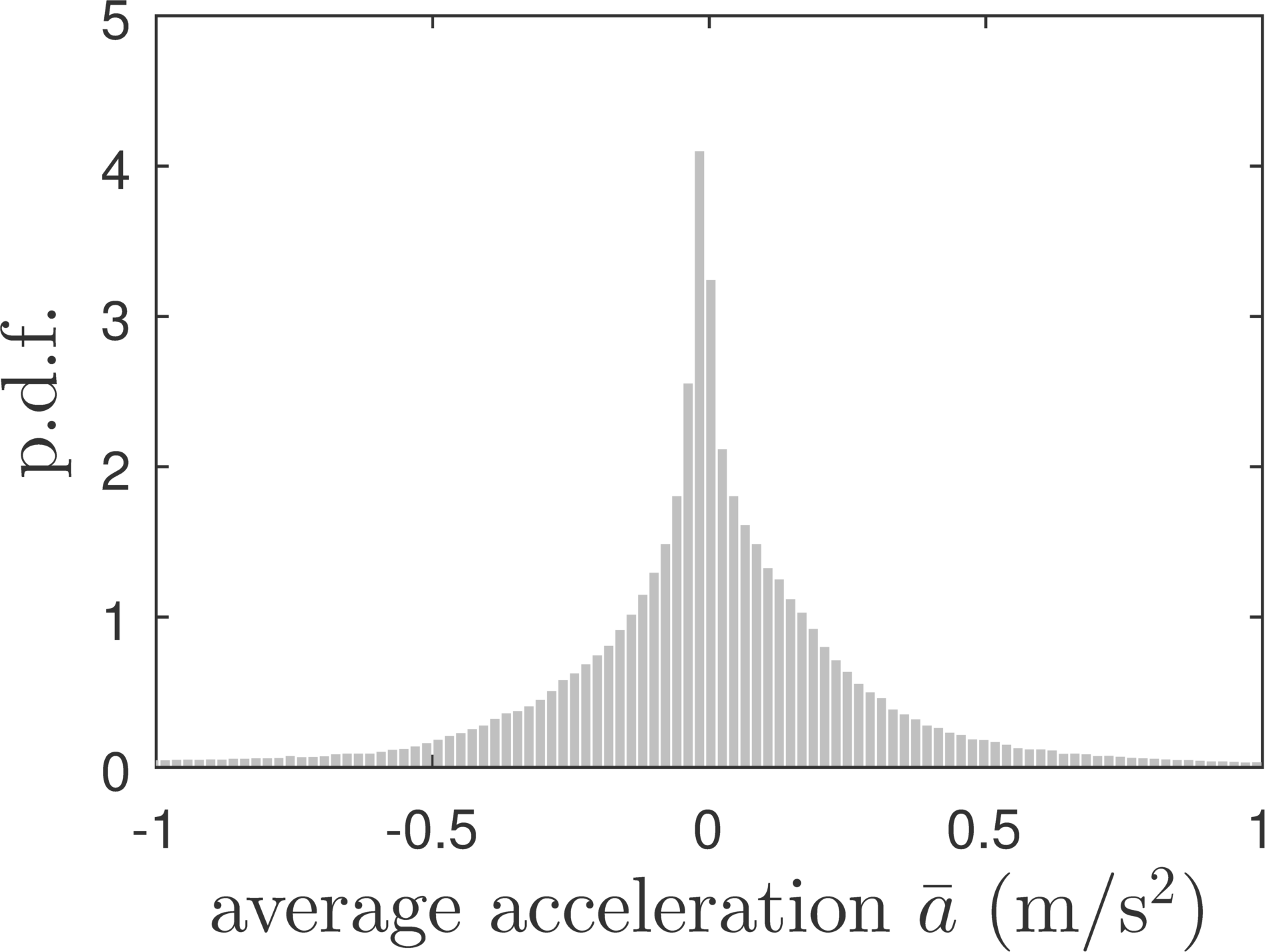}&
		\includegraphics[width=0.47\linewidth]{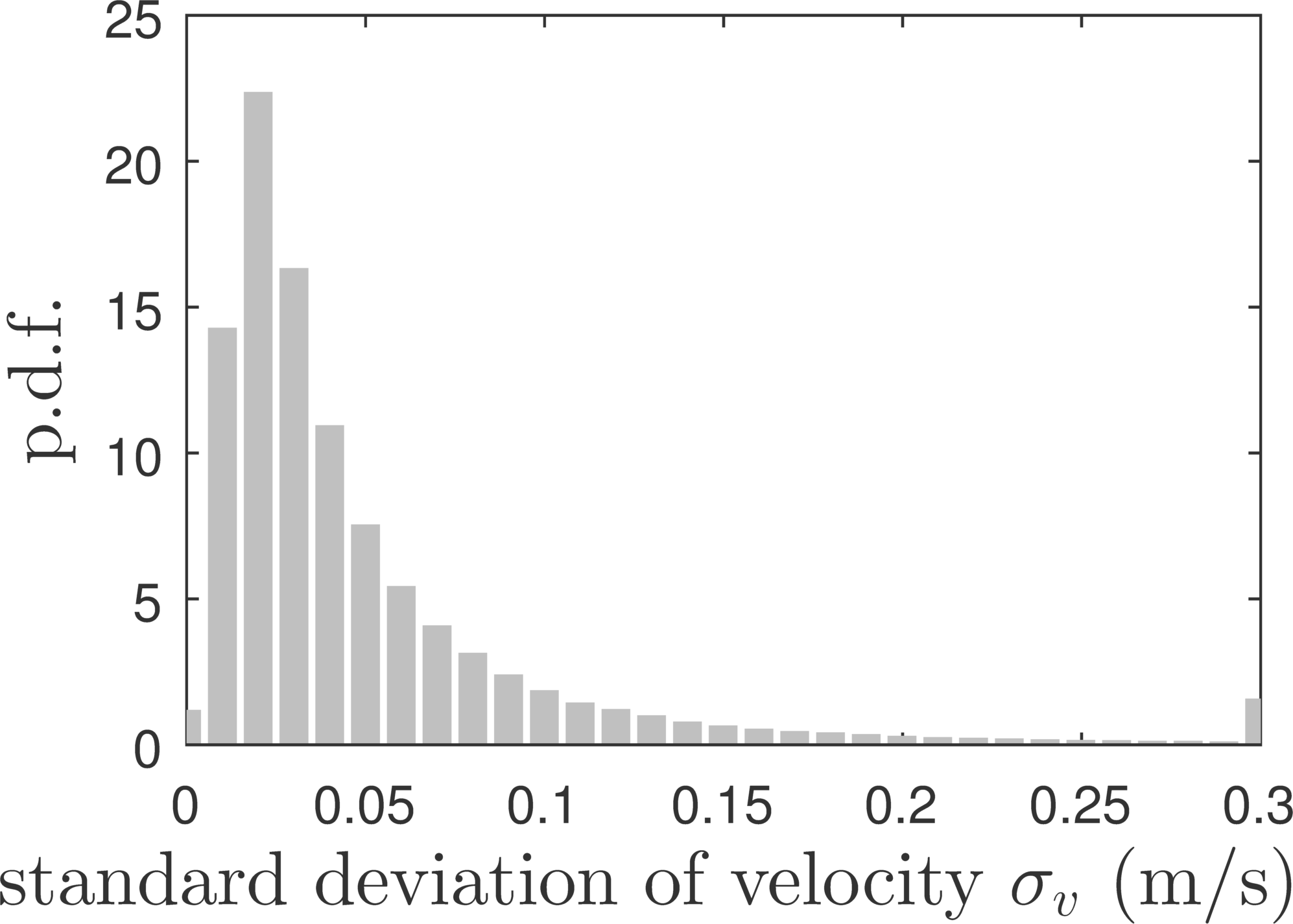}\\
		\begin{small} (e) \end{small} &\begin{small} (f) \end{small} \\
		\includegraphics[width=0.47\linewidth]{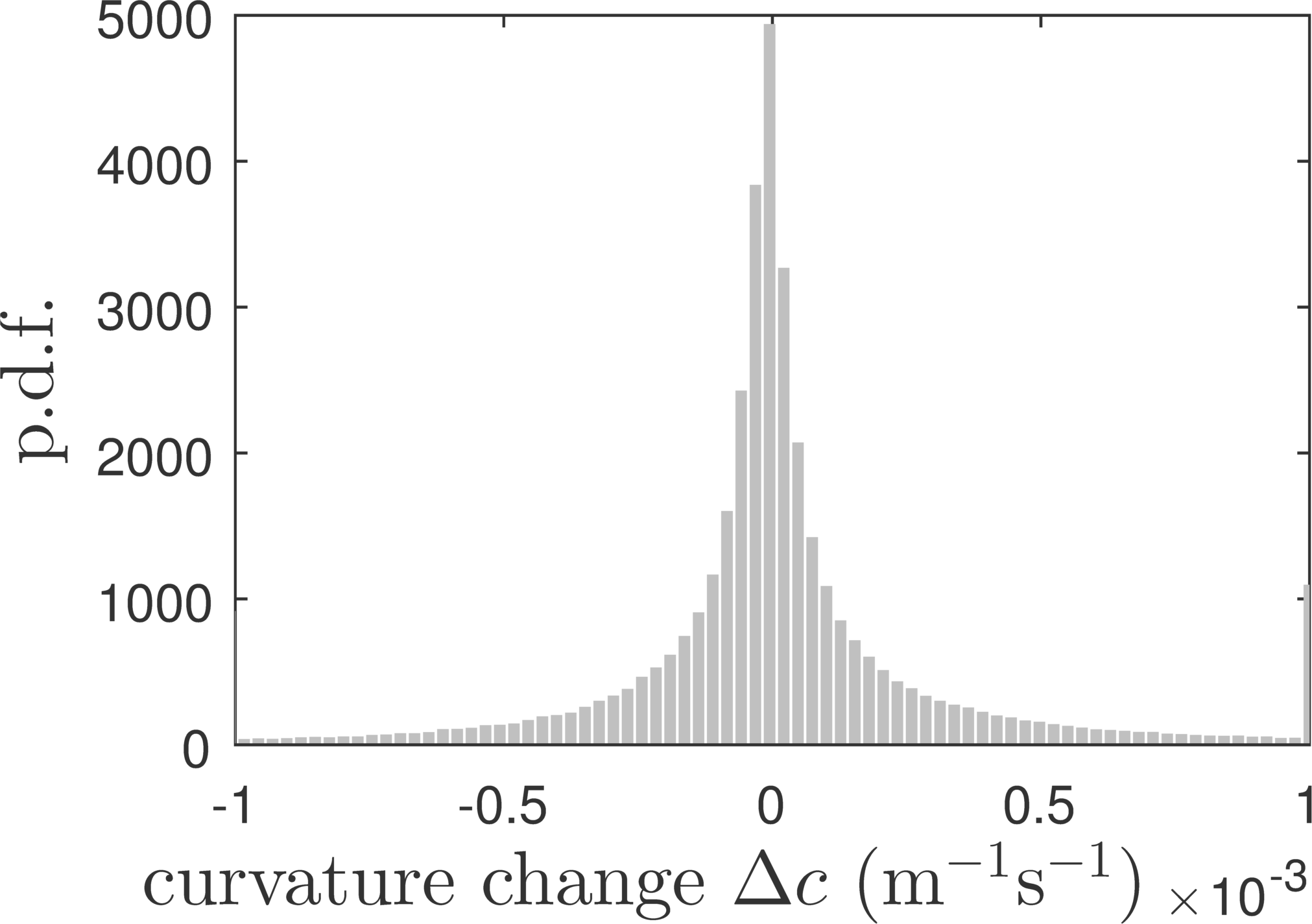}&
		\includegraphics[width=0.47\linewidth]{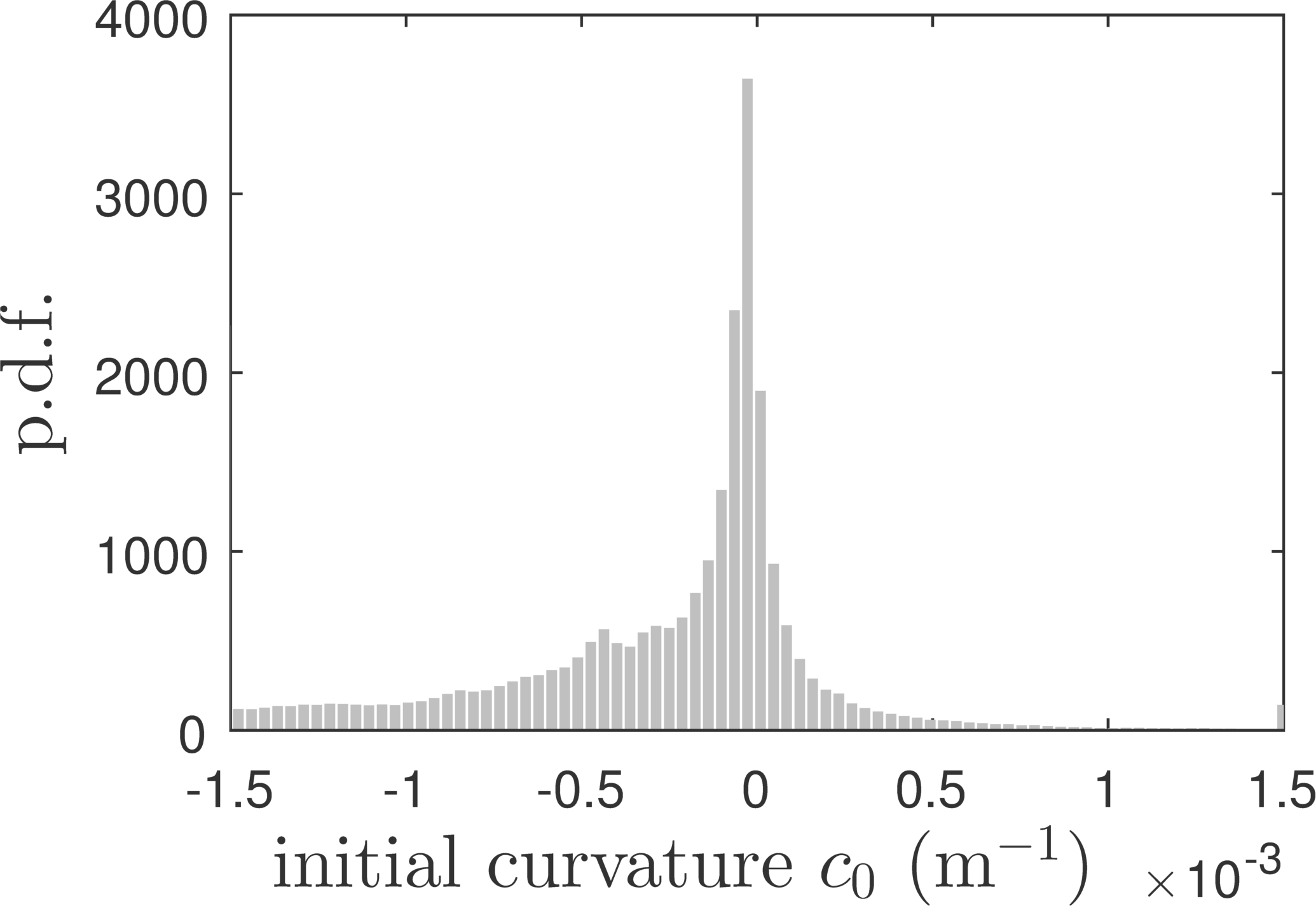}\\
		\begin{small} (g) \end{small} &\begin{small} (h) \end{small} 
	\end{tabular}
	\caption{Marginal distributions of left lane departure variables}
	\label{fig:marginal_L}
\end{figure}

\begin{figure}[t]
	\centering
	\begin{tabular}{cc}
		\includegraphics[width=0.47\linewidth]{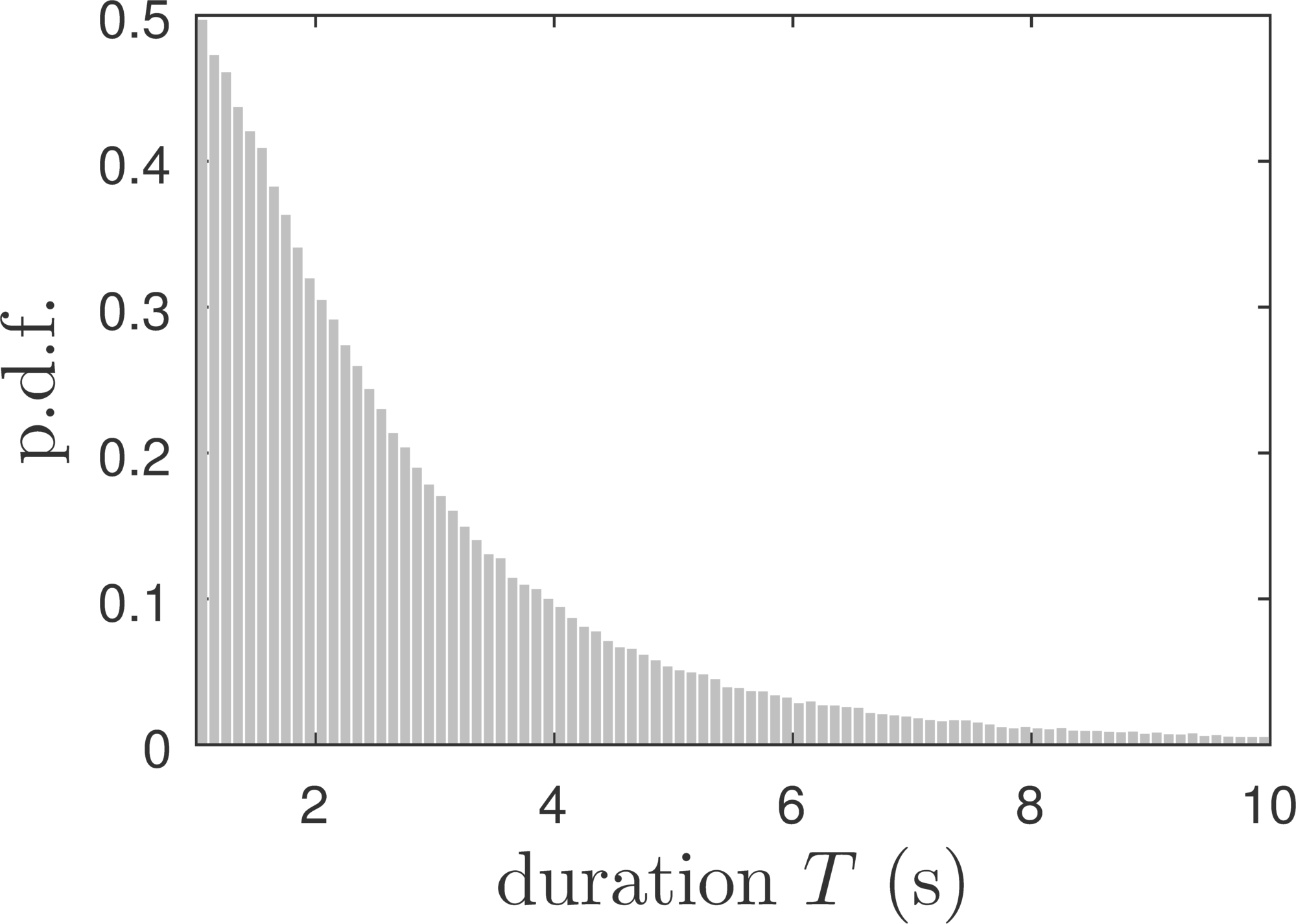} &
		\includegraphics[width=0.47\linewidth]{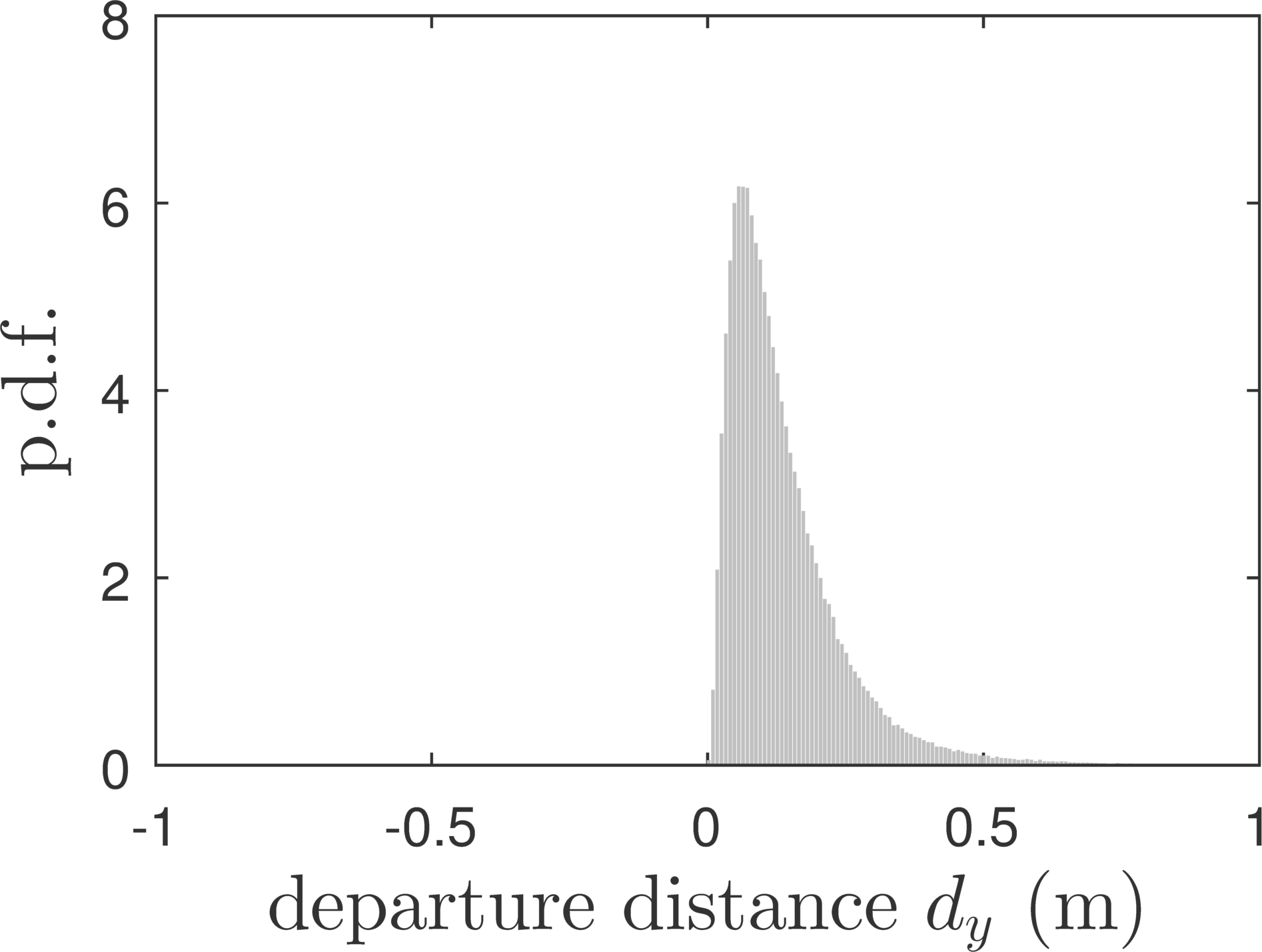}\\
		\begin{small} (a) \end{small} &\begin{small} (b) \end{small} \\
		\includegraphics[width=0.47\linewidth]{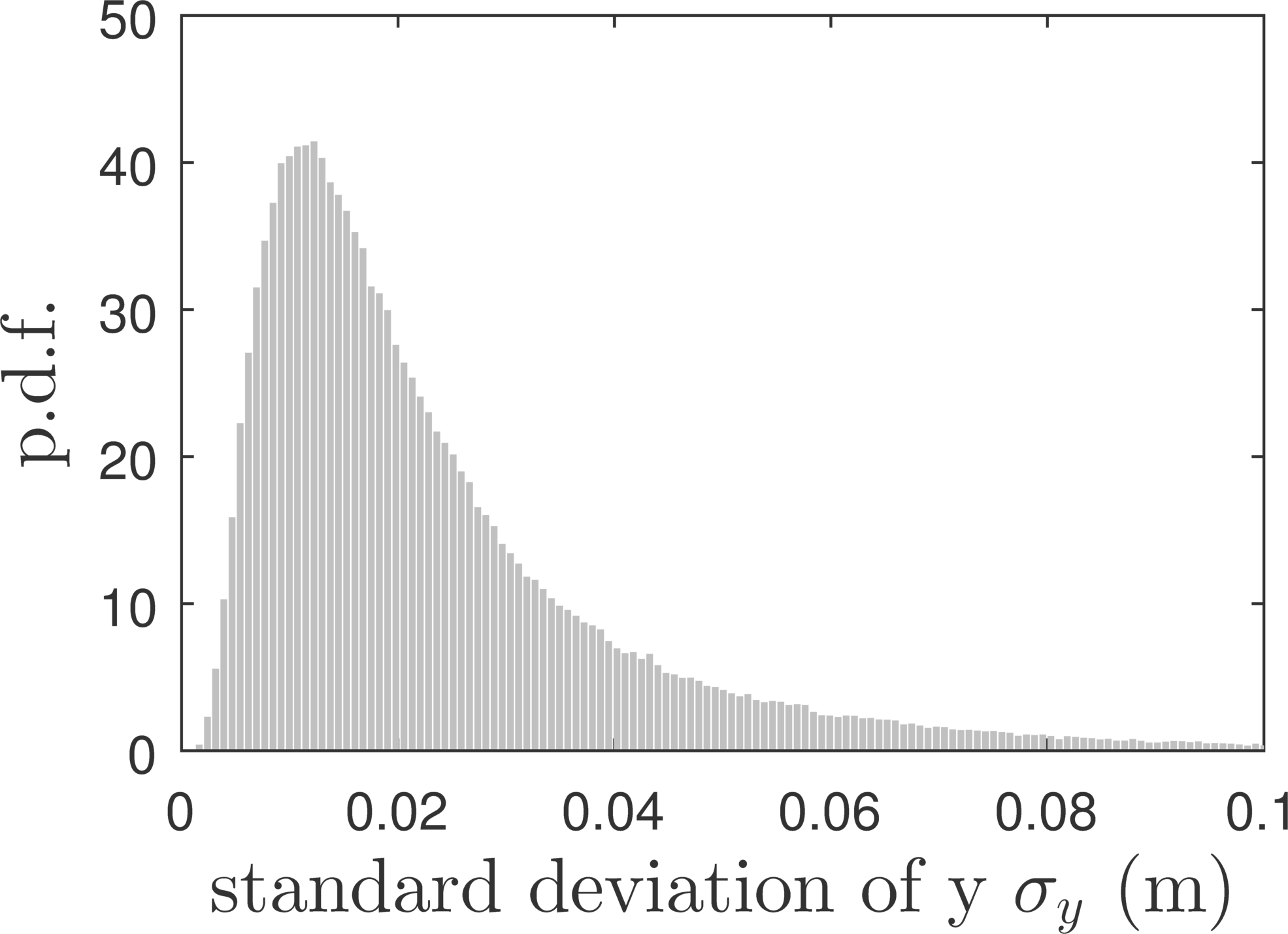}&
		\includegraphics[width=0.47\linewidth]{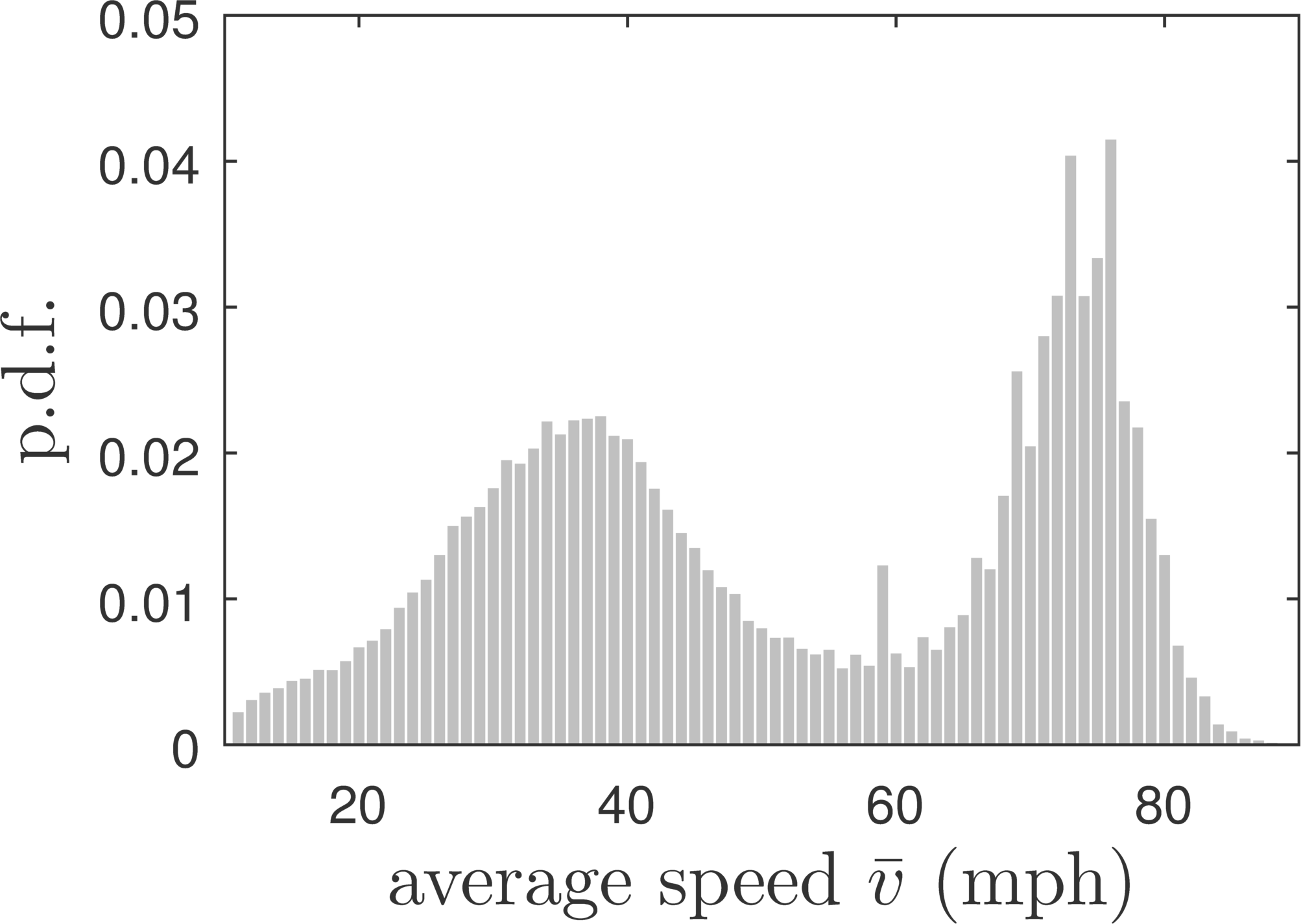}\\
		\begin{small} (c) \end{small} &\begin{small} (d) \end{small} \\
		\includegraphics[width=0.47\linewidth]{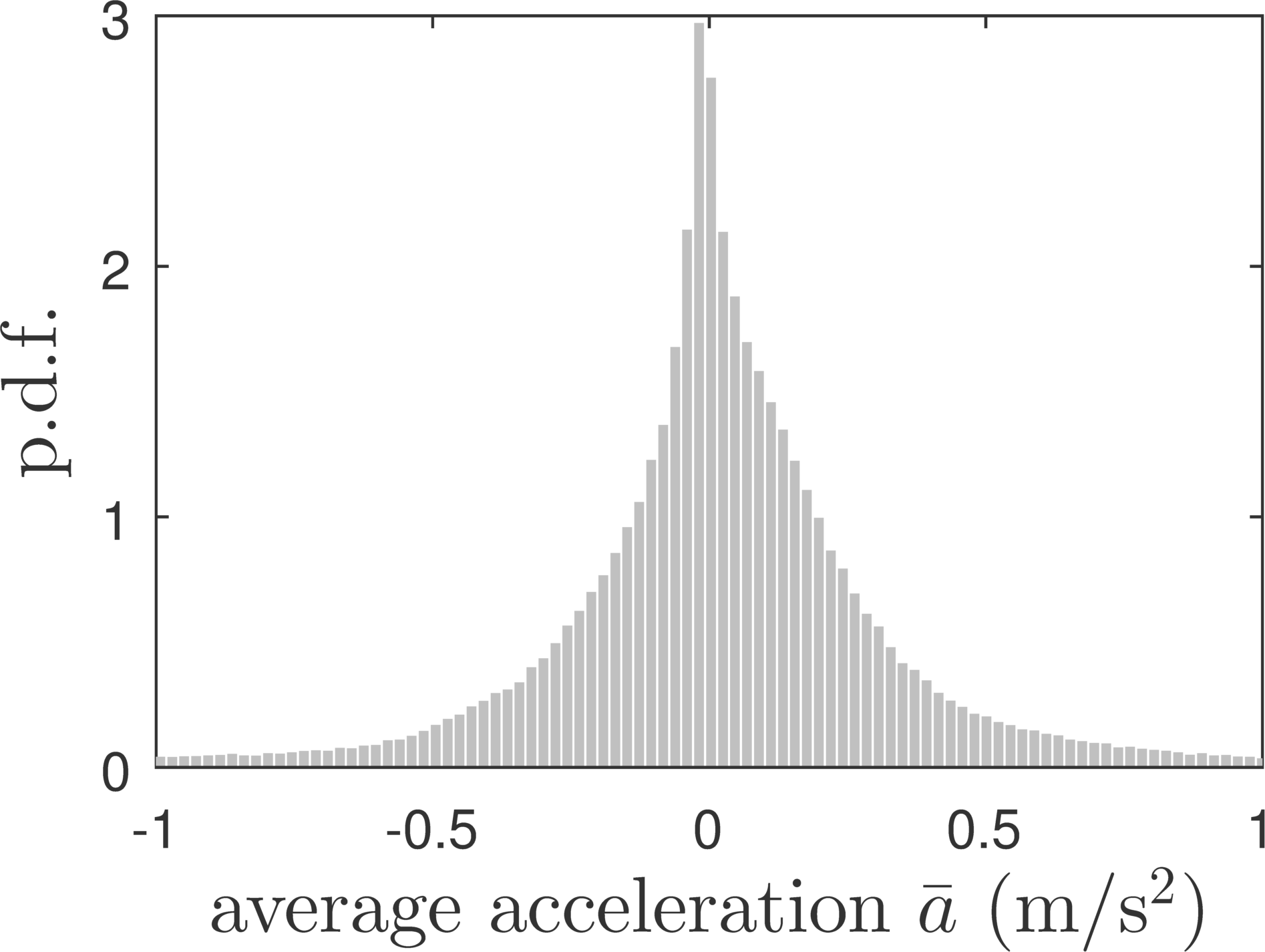}&
		\includegraphics[width=0.47\linewidth]{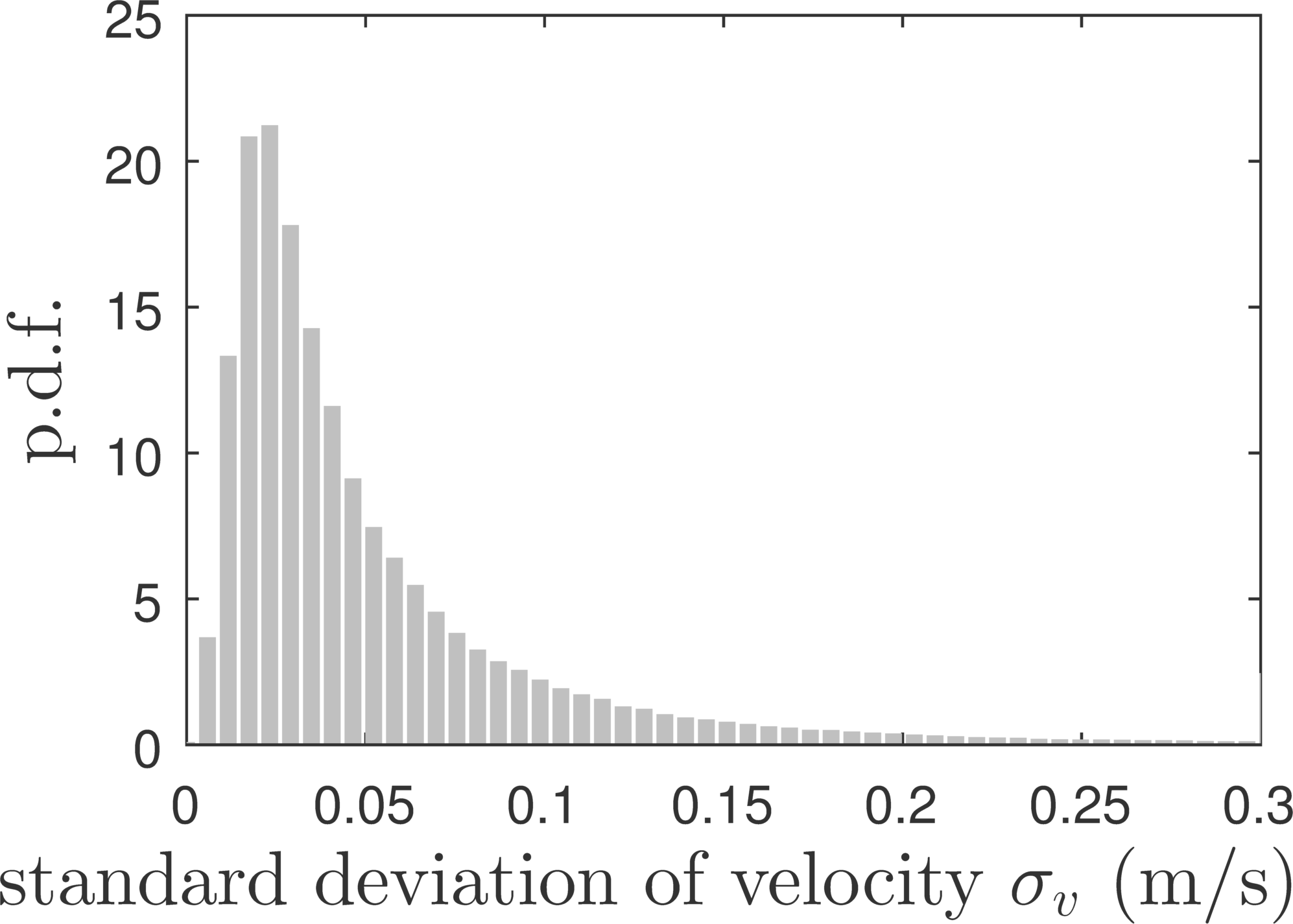}\\
		\begin{small} (e) \end{small} &\begin{small} (f) \end{small} \\
		\includegraphics[width=0.47\linewidth]{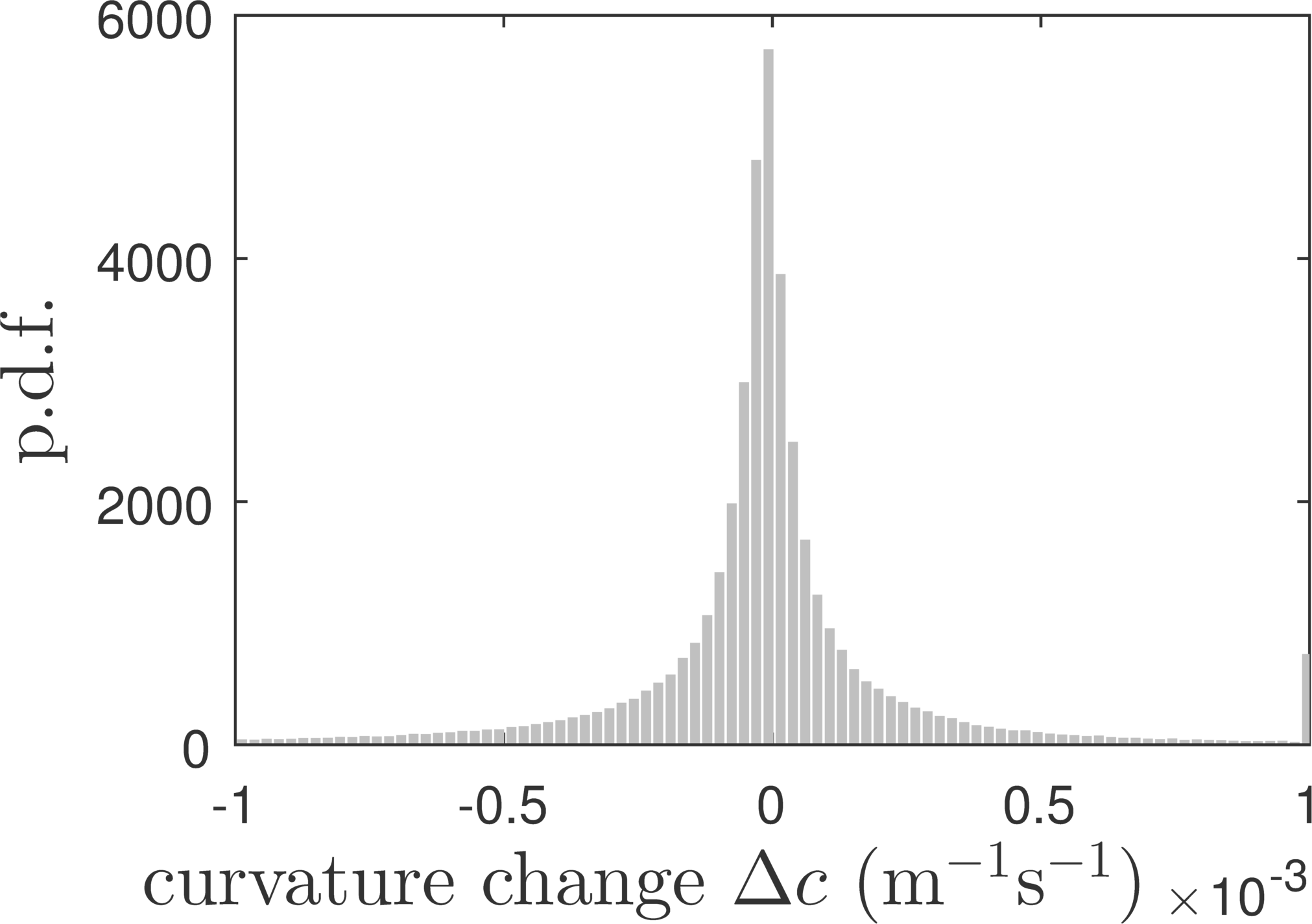}&
		\includegraphics[width=0.47\linewidth]{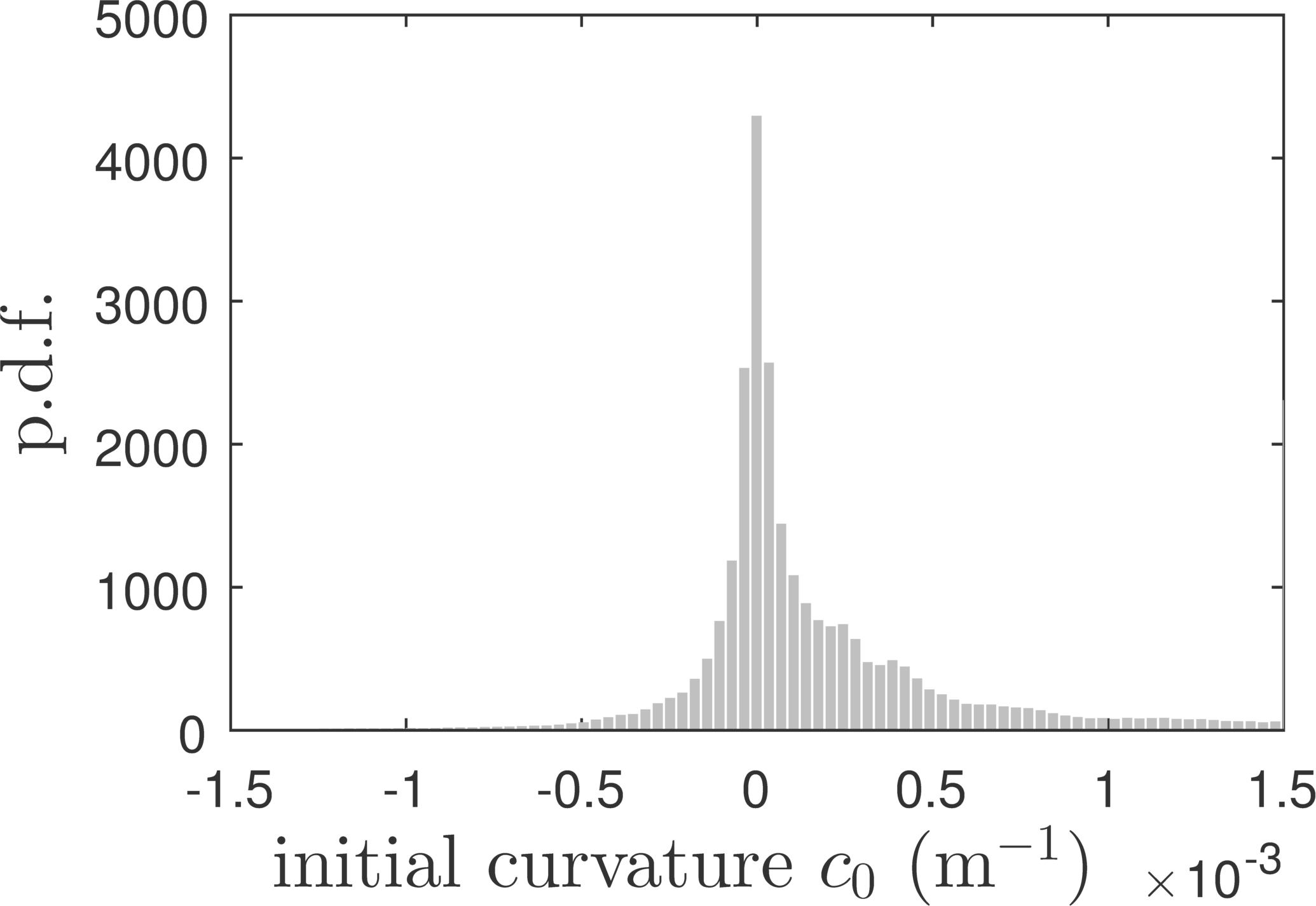}\\
		\begin{small} (g) \end{small} &\begin{small} (h) \end{small} 
	\end{tabular}
	\caption{Marginal distributions of right lane departure variables}
	\label{fig:marginal_R}
\end{figure}

Velocity is approximated using a linear function of time.
\begin{equation}
v(t)=\tilde{v}(t)+\epsilon_v(t)=\bar{a}(t-T/2)+\bar{v} +\epsilon_v(t)
\end{equation}
where $\bar{v}=d_x/T$  is the average speed. The average acceleration $\bar{a}$ is calculated from the least square method.
Similarly, human uncertainties on velocity can be calculated from 
\begin{equation}
\label{eq:sigma_v}
\sigma_v = \sqrt{\dfrac{1}{L-1}\sum_{l=0}^{L}|\epsilon_v(l)-\bar{\epsilon}_v|^2}
\end{equation}
 
Lane curvature is also modeled as a linear function of time:
\begin{equation}
c(t) = \tilde{c}(t)+\epsilon_{c}=\frac{\Delta c}{T}t+c_{0}+\epsilon_{c}
\end{equation}
where $c_{0}$ is the initial curvature and $\Delta c$ is the change in the curvature. To smooth the curvature data, we use linear regression to estimate  $c_{0}$ and $\Delta c$ such that $\sum_{l=1}^L |c(l)-\tilde{c}(l)|^2 $ is minimized.
The benefits of this model are that it reduces the dimensions of the original data while capturing the stochastic variance in human driving  with parameters that have physical meanings.

\section{Statistical Model Fitting}
\begin{figure}[t]
	\centering
	\includegraphics[width=\linewidth]{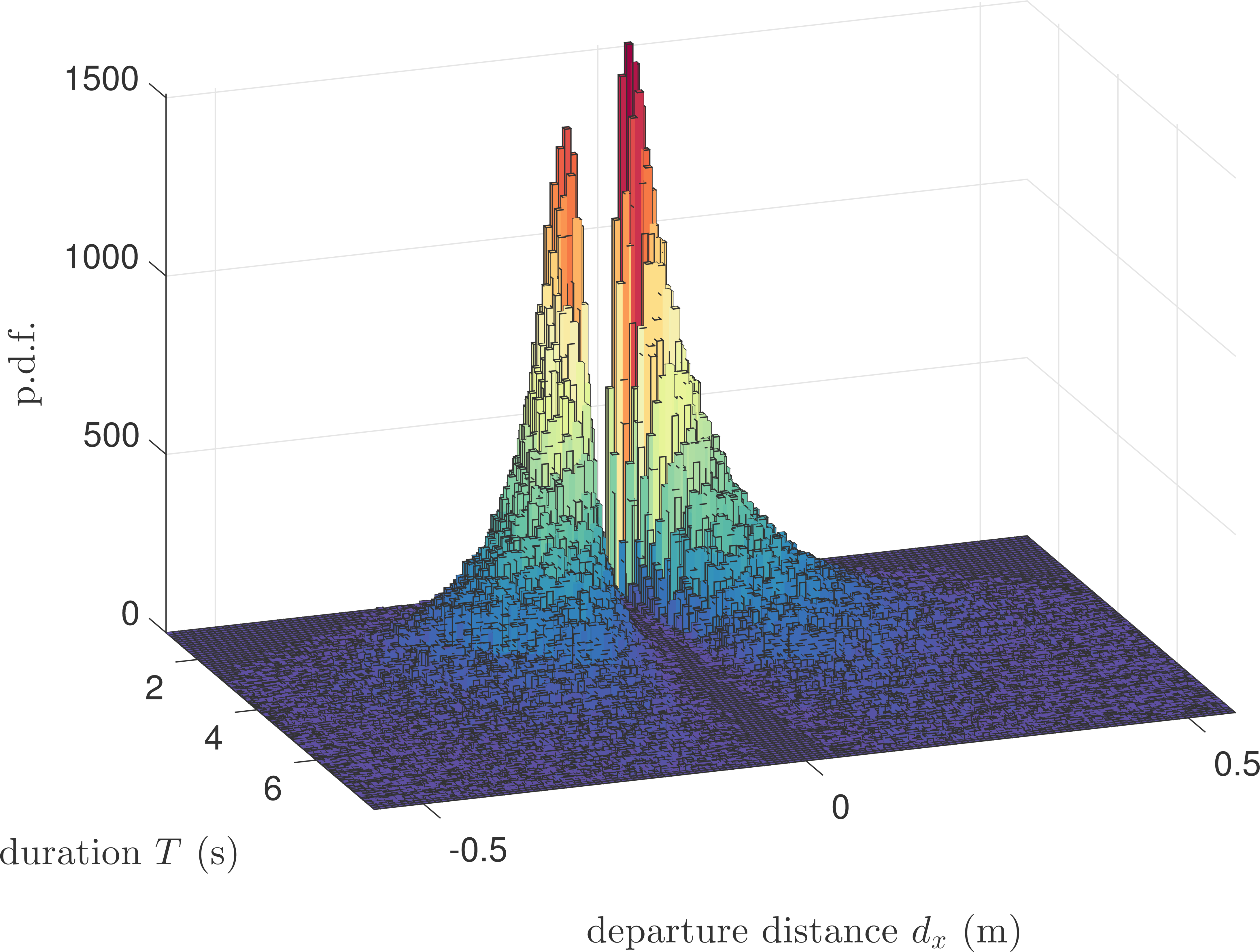}
	\caption{Joint distributions of duration and maximum departure distance}
	\label{fig:joint}
\end{figure}

In this section, we develop a statistical model to describe the joint distribution of the seven parameters identified in previous section. Let $\bm{\xi} ^{(n)}=[T^{(n)}, d_y^{(n)}, \sigma_y^{(n)}, \bar{v}^{(n)}, \bar{a}^{(n)}, \sigma_v^{(n)}, c_{0}^{(n)}, \Delta c^{(n)}] $ where $ n=1,2,...,N $ is the index of departure events. The marginal distribution of $\bm{\xi}$ for left lane departure events and right lane departure events are shown Fig. \ref{fig:marginal_L} and Fig. \ref{fig:marginal_R}, respectively. We can see that each variable follows a different distribution. From the joint distribution shown in Fig. \ref{fig:joint}, a clear dependence between variables can be seen. A flexible probability density function ({\it p.d.f.}) that can model the multi-variate distribution is needed.

In light of its flexibility and ease of training, the Gaussian Mixture Model (GMM) has been widely used in applications such as speech recognition \cite{jonas2003bounded}, pattern recognition \cite{nguyen2014bounded}, and driving behaviors \cite{Zhao2015AcceleratedData}. In this research, we use the BGM model to take the physical boundaries into account. The BGM model is valuable because it considers the variable boundaries while preserving a tractable form when using the EM algorithm to train the model.

The probability density function of a BGM model can be expressed as
\begin{equation}
f_{\mathcal{BM}}(\bm{\xi})=\frac{f_{\mathcal{GM}}(\bm{\xi})}{\int_{\bm{b_l}}^{\bm{b_u}}f_{\mathcal{GM}}(\bm{u})d\bm{u}}
\end{equation}
where $ f_{\mathcal{GM}}(\bm{\xi}) $ is a normal GMM, i.e.
\begin{equation}
f_{\mathcal{GM}}(\bm{\xi}|\bm{\Theta})=\sum_{k=1}^{K}\pi_k g_k(\bm{\xi};\bm{\theta}_k)
\end{equation}
where $ \pi_k \in [0,1] $ are mixing weights with $ \sum_k \pi_k = 1 $, $ g_k $ is the $ k^{\textrm{th}}$ d-dimensional Gaussian distribution component parameterized by $ \bm{\theta_k}=[\bm{\mu_k},\bm{\sigma_k}] $. Here we assume that the boundary is a hyper-rectangle in $ \mathbb{R}^d $ with two vertices $ \bm{b_u} =[b^u_{1},...,b^u_{d}]^T $ and $ \bm{b_l} =[b^l_{1},...,b^l_{d}]^T$ on the diagonal opposites such that $ \bm{b_l}<\bm{\xi}^{(n)}<\bm{b_u} $, $ \bm{\Theta}=[\pi_1,...,\pi_K,\bm{\theta_1},...,\bm{\theta_K}] $.

It can be derived that $ f_{\mathcal{BM}} $ is also a mixture
\begin{equation}
f_{\mathcal{BM}}=\sum_{k=1}^{K}\eta_k f_k(\bm{\xi})
\end{equation}
with mixing weights $ \eta_k $ and component density functions $ f_k $:
\begin{align}
	\eta_k&=\pi_k\frac{\int_{\bm{b_l}}^{\bm{b_u}}g_k(\bm{u})d\bm{u}}
	{\int_{\bm{b_l}}^{\bm{b_u}}f_{\mathcal{GM}}(\bm{u})d\bm{u}}\\
	f_k(\bm{\xi})&=\frac{g_k(\bm{\xi})}
	{\int_{\bm{b_l}}^{\bm{b_u}}f_{\mathcal{GM}}(\bm{u})d\bm{u}}
\end{align}

\begin{figure}[t]
	\centering
	\begin{subfigure}{0.48\textwidth}
	\includegraphics[width=\linewidth]{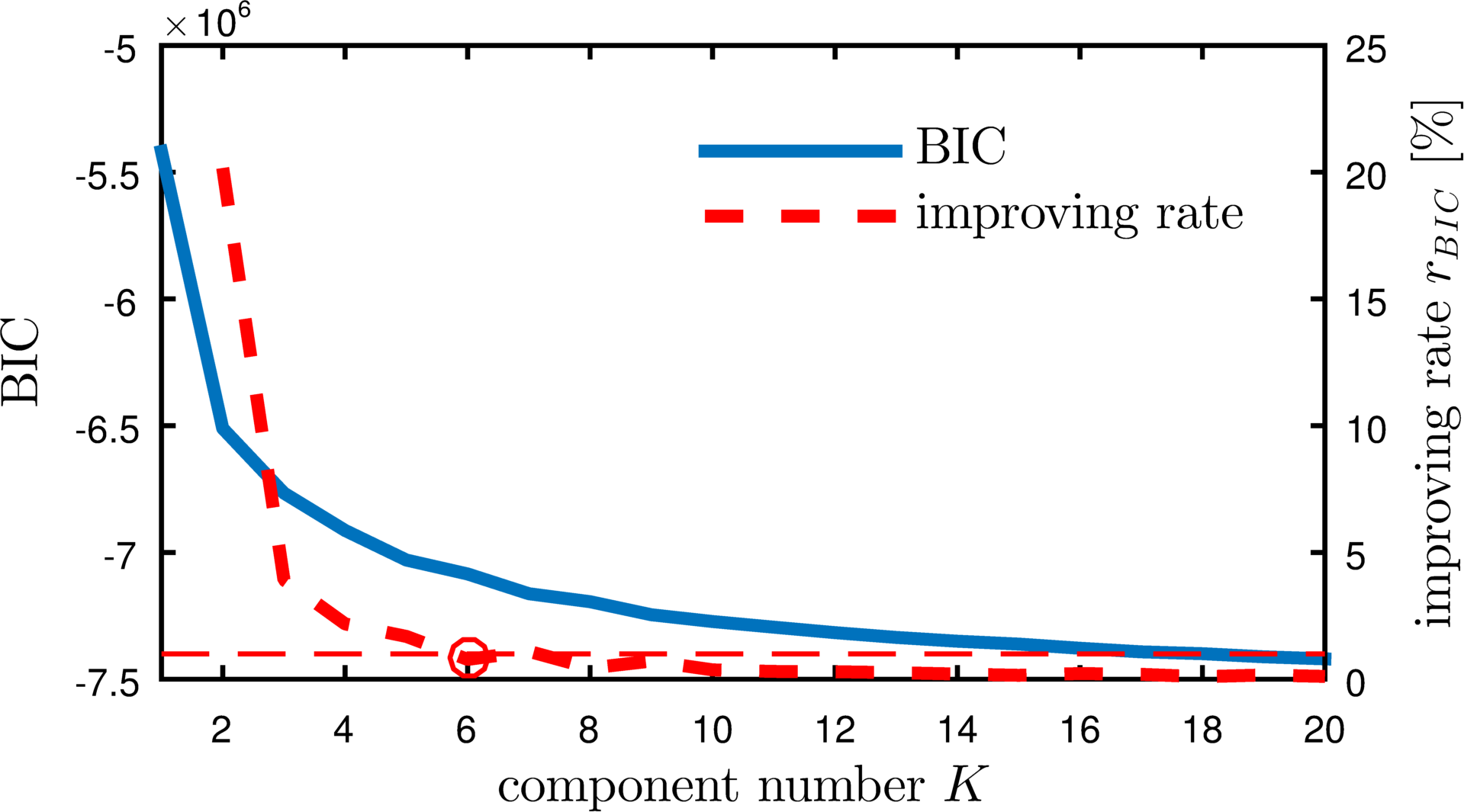}
	\caption{Left departure}
	\end{subfigure}
	\begin{subfigure}{0.48\textwidth}
		\includegraphics[width=\linewidth]{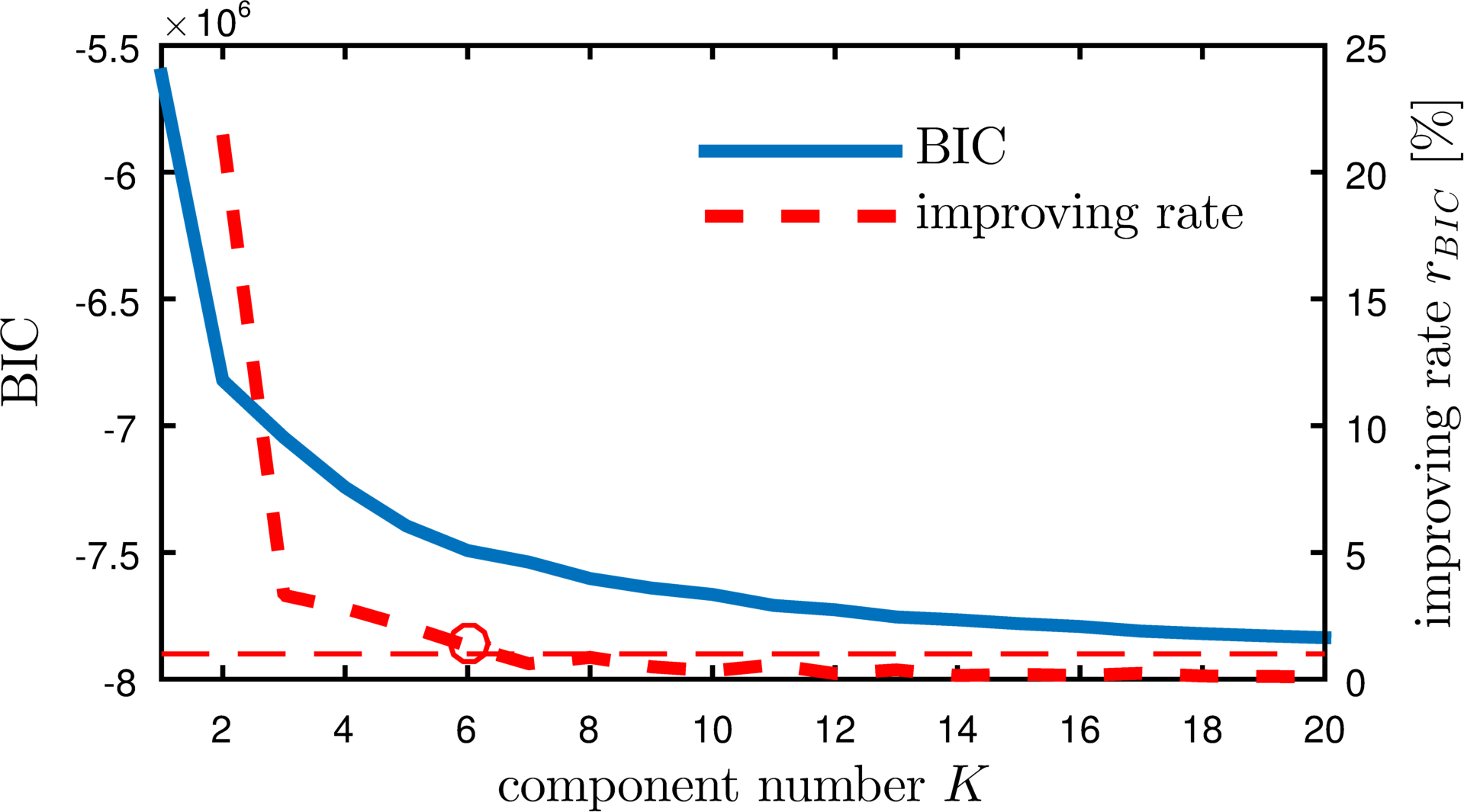}
		\caption{Right departure}
	\end{subfigure}
	\caption{BIC with different numbers of  BGM components.}
	\label{fig:BIC}
\end{figure}

The log-likelihood function of $f_{\mathcal{BM}}$ can be expressed as
\begin{align}
\mathcal{L}_\mathcal{B}(\bm{\Theta}) &= \ln\prod_{n}\sum_{k}z_k^n \eta_k f_k(\bm{\xi^n})\\
&=
\sum_{n} \sum_{k}
z_k^n[\ln \eta_k+\ln f_k(\bm{\xi^n})-
\ln \int_{\bm{b_l}}^{\bm{b_u}}f_k(\bm{u})d\bm{u}]
\end{align}
The expectation of $ \mathcal{L}_\mathcal{B}(\bm{\Theta}) $ can be calculated from
\begin{align}
&\mathcal{Q}_\mathcal{B}(\bm{\Theta^{(i+1)}};\bm{\Theta^{(i)}})=
\mathbb{E}[\mathcal{L}_\mathcal{B}(\bm{\Theta})|\bm{\xi^{1:N}};\bm{\Theta^{(i)}}]\\
&= \sum_{n} \sum_{k}
\langle z_k^n \rangle[\ln \eta_k+\ln f_k(\bm{\xi^n})-
\ln \int_{\bm{b_l}}^{\bm{b_u}}f_k(\bm{u})d\bm{u}]
\end{align}
where the latent variable 
\begin{equation}
\langle z_k^n \rangle:=\mathbb{P}(z_k^n=1|\bm{\xi^n})
\end{equation}
As shown in \cite{Lee2012EMData}, the EM iteration can be calculated from
\begin{align}
\eta_k&=\frac{1}{N}\sum_n \langle z_k^n \rangle\\
\bm{\mu_k}&=\frac{\sum_n \langle z_k^n \rangle \bm{\xi^n}}
	{\sum_n \langle z_k^n \rangle}-\bm{m_k}\\
\bm{\Sigma_k}&=\frac{\sum_n \langle z_k^n \rangle (\bm{\xi^n}-\bm{\mu_k})}
	{\sum_n \langle z_k^n \rangle}+H_k
\end{align}
where
\begin{align}
\bm{m_k} &= \mathcal{M}^1(0,\Sigma_k;[\bm{b_l}-\bm{\mu_k},\bm{b_u}-\bm{\mu_k}])\\
H_k &= \Sigma_k-\mathcal{M}^2(0,\Sigma_k;[\bm{b_l}-\bm{\mu_k},\bm{b_u}-\bm{\mu_k}])
\end{align}
with $\mathcal{M}^1$ and $\mathcal{M}^2$ represent the first order and second order moment generated function of $f_{\mathcal{BM}}$.

The component number of the BGM model is chosen based on a numerical analysis of the Bayesian Information Criterion (BIC) \cite{Box2015TimeControl} . As shown in Fig. \ref{fig:BIC}, the BIC decreases very slowly and begins to oscillate when the number of components is greater than 10 for both left and right lane departure cases. Therefore we chose $K=10$. 






\section{Evaluation  of a Semi-Autonomous Lane Departure Correction System}
An LDC system is designed to demonstrate the evaluation approach. First, we describe the vehicle dynamics during the lane departure. Then we introduce an LDC system by controlling the steering wheel.
\subsection{The Vehicle Dynamics}

\begin{figure}[t]
	\centering
	\includegraphics[width=0.9\linewidth]{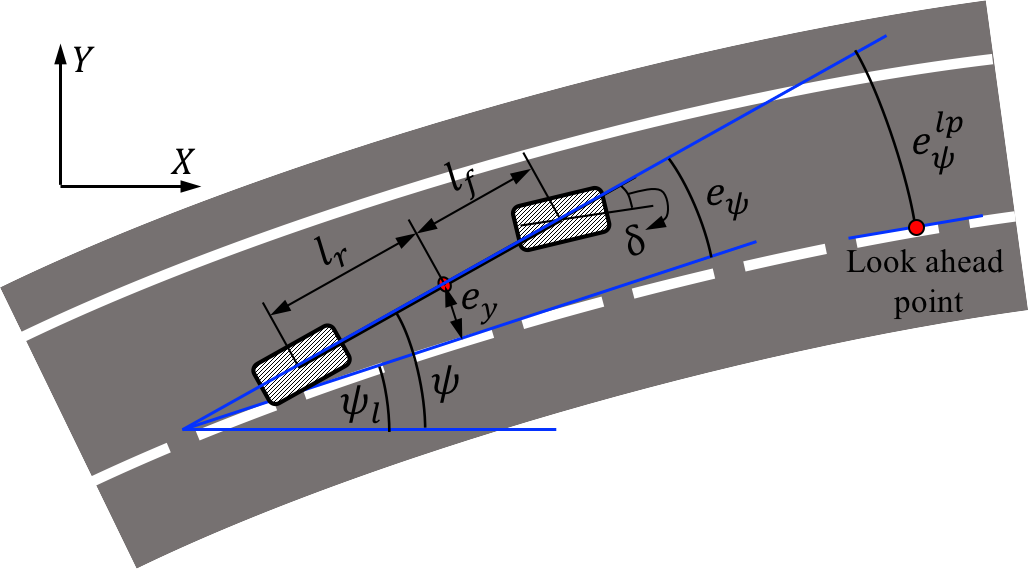}
	\caption{Bicycle model and the related variables in a curve path.}
	\label{fig:vehDyn}
\end{figure}

When vehicle speed varies slightly and the slip values are small, as at the beginning of lane departure drifting events, a linear vehicle model can be employed. As shown in Fig. \ref{fig:vehDyn}, a simplified vehicle model is used, where two front wheels and two rear wheels are lumped together, respectively. $\psi$ is the heading angle and $\psi_{l}$ is the tangent direction of the lane. We define the heading error $ e_\psi $ and offset error $ e_y $ as 
\begin{equation}
\begin{split}
e_\psi & =\psi-\psi_l \\
e_y & =y-\frac{w_v}{2}+\frac{w_{l}}{2}
\end{split}
\end{equation}
where $w_v$ and $w_{l}$ are vehicle width and lane width, respectively.
The vehicle dynamic can be expressed in a state space form
\begin{equation}
\label{eq:VehivleDynamic}
\dot{\bm{x}}(t)=\bm{A}\bm{x}(t)+\bm{B}\delta(t)+\bm{E}\dot{\psi}_l(t),
\end{equation}
\begin{gather*}
\resizebox{\hsize}{!}{$
\mathbf{A} = 
\begin{bmatrix}[1.5]
0 & 1 & 0 & 0\\
0 & -\frac{2 C_{\alpha f}+2 C_{\alpha r}}{M v_x} & \frac{2 C_{\alpha f}+2 C_{\alpha r}}{M} 
													& -\frac{2 C_{\alpha f} l_f+2 C_{\alpha r}l_r}{M v_x}\\
0 & 0 & 0 & 1\\
0 & -\frac{2 C_{\alpha f} l_f-2 C_{\alpha r}l_r}{I_z v_x} & \frac{2 C_{\alpha f} l_f-2 C_{\alpha r}l_r}{I_z} 
													& - \frac{2 C_{\alpha f} l^2_f+2 C_{\alpha r}l^2_r}{I_z v_x}
\end{bmatrix} $}\\
\bm{B} = 
\begin{bmatrix}[1.5]
0\\
\frac{2 C_{\alpha f}}{m}\\
0\\
\frac{2 C_{\alpha f}l_f}{I_z}
\end{bmatrix}
,\qquad 
\bm{E}= 
\begin{bmatrix}[1.5]
0\\
-\frac{2 C_{\alpha f}l_f-2 C_{\alpha r}l_r}{m v_x}-v_x\\
0\\
-\frac{2 C_{\alpha f}l^2_f+2 C_{\alpha r} l^2_r}{I_z v_x}
\end{bmatrix}
\end{gather*}
where $\bm{x}(t)=[e_y,\dot{e}_y,e_\psi ,\dot{e}_\psi]^{T}\in \mathbb{R}^{4\times 1},\ \bm{A}\in \mathbb{R}^{4\times 4},\ \bm{B}\in \mathbb{R}^{4\times 1},\ \bm{E}\in \mathbb{R}^{4\times 1}$, $C_{\alpha f}$ and $C_{\alpha r}$  are tire stiffness, $l_f$ and $l_r$ are the longitudinal distance from the center of gravity to the front axle and rear axle, $m$ is the total mass, $I_z$  is the inertia of $z$ axis, and $\delta$ is the average of the steering angles of the front two wheels.



\subsection{Controller Design}
Define the preview orientation error (Fig. \ref{fig:vehDyn})
\begin{equation}\label{eq:25}
e_\psi^{lp} = \psi-\psi_l^{lp}
\end{equation}
Let 
\begin{equation}\label{eq:26}
\Delta\psi_l=\psi_l-\psi_l^{lp}
\end{equation}
Substituting (\ref{eq:26}) to (\ref{eq:25}), we have
\begin{equation}
e_\psi^{lp} = \psi-(\psi_l-\Delta\psi_l) = e_\psi+\Delta\psi_l
\end{equation}
Consider the control law
\begin{equation}
\delta = K_y e_y+K_\psi e_\psi^{lp}
\end{equation}
Letting $\bm{F} = [K_y,0,K_\psi,0]$ and $G=K_\psi$, we have
\begin{equation}
\label{eq:ctrl}
\begin{split} 
\delta & = K_y e_y+K_\psi e_\psi+K_\psi\Delta\psi_l\\
& = \bm{F}\bm{x}+G\Delta\psi_l
\end{split}
\end{equation}
The semi-autonomous system kicks in when a certain threshold is reached. In this paper we will evaluate a design with a trigger  $y>y_s$, where $ y_s $ is the predefined threshold.

Based on the above discussion, we rewrite (\ref{eq:VehivleDynamic}) as follows
\begin{equation}\label{eq:VehivleDynamic2}
\dot{\bm{x}}(t)=\bm{A_c}\bm{x}(t)+\bm{B_c} \bm{l}(t),
\end{equation}
where $\bm{A_c}=\bm{A}+\bm{B}\bm{F}$, $\bm{B_c}=[\bm{E},\bm{B}G]$, $ \bm{l}(t)=[\dot{\psi}_l(t), \Delta\psi_l(t)]^T$. Substituting  (\ref{eq:ctrl}) into (\ref{eq:VehivleDynamic2}), we get the closed loop form
\begin{equation}
\dot{\psi}_l(t)=v_x(t_s)c(t)=\frac{v_x(t_s) \Delta c}{T}t+v_x(t_s)c_0
\end{equation}

\begin{equation}
\Delta\psi_l(t)=\int_{t} ^{t+T_{lp}} \dot{\psi}_{l}(t)  dt
=A_{\Delta\psi_l} t+B_{\Delta\psi_l}
\end{equation}
where $A_{\Delta\psi_l} = \frac{\Delta c \cdot T_{lp} \cdot v_x(t_s)}{T}$ and $B_{\Delta\psi_l}=\frac{\Delta c\cdot T_{lp}^2 \cdot v_x(t_s)}{2 T}+v_x(t_s)\cdot c_0\cdot T_{lp}$. The initial condition of the close loop control is $e_y(t_s)=y(t_s)+(w_l-w_v)/2$ and $e_\psi(t_s) = \arctan(\frac{v_y(t)}{v_x(t)})$, where $v_y(t_s) = \frac{d e_y(t)}{dt}|_{t = t_s} $, $ t_s $ is the time of start point of lane departure. Table \ref{tab:simpara} provides the parameters used in the simulation.

\begin{figure}[t]
	\centering
	\includegraphics[scale = 0.45]{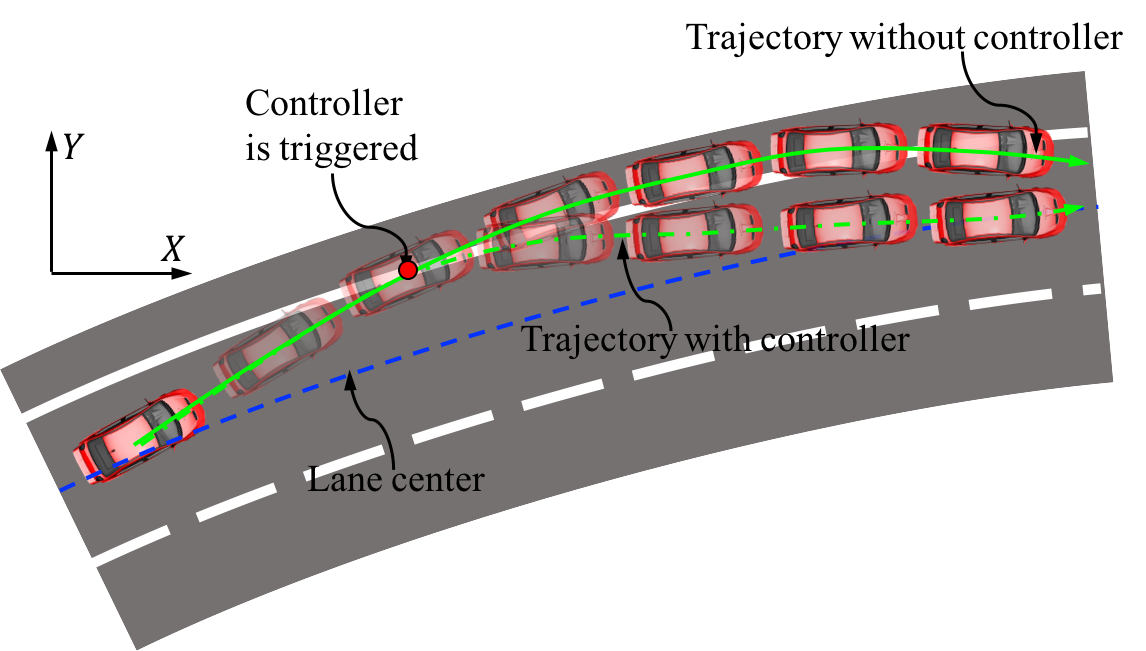}
	\caption{Illustrations of comparison between trajectories with and without the controller.}
	\label{Illus_ex}
\end{figure}

\begin{table}[htbp]
	\caption{Simulation Parameters}
	\label{tab:simpara}
	\centering
	\begin{tabular}{c c c|c c c}
		\hline\hline
		Var & Unit & Value & Var & Unit & Value\\
		\hline
		$C_{\alpha f}$ &N/rad	&80000	&$l_{r}$	&m			&1.47	\\
		$C_{\alpha r}$	&N/rad	&80000	&$I_z$		&$kgm^2$	&3344	\\
		$l_{f}$			&m		&1.43	&M			&kg			&1000	\\
		$T_lp$			&s		&2		&$D_y$		&m			&0.5	\\
		$T_s$		&s		&0.05 &$w_l$		&m		& 3.6 \\
		$K_y$		&rad/m	&-0.005 & $ w_v $ & m & 1.9  \\
		$K_\psi$	&rad/rad&-0.2 & $ y_s $ & m & 0.2\\
		\hline
		\hline
	\end{tabular}
\end{table}

\subsection{Simulation Results}

\begin{figure}[t]
	\centering
	\begin{subfigure}[t]{0.5 \textwidth}
		\centering
		\includegraphics[scale = 0.56]{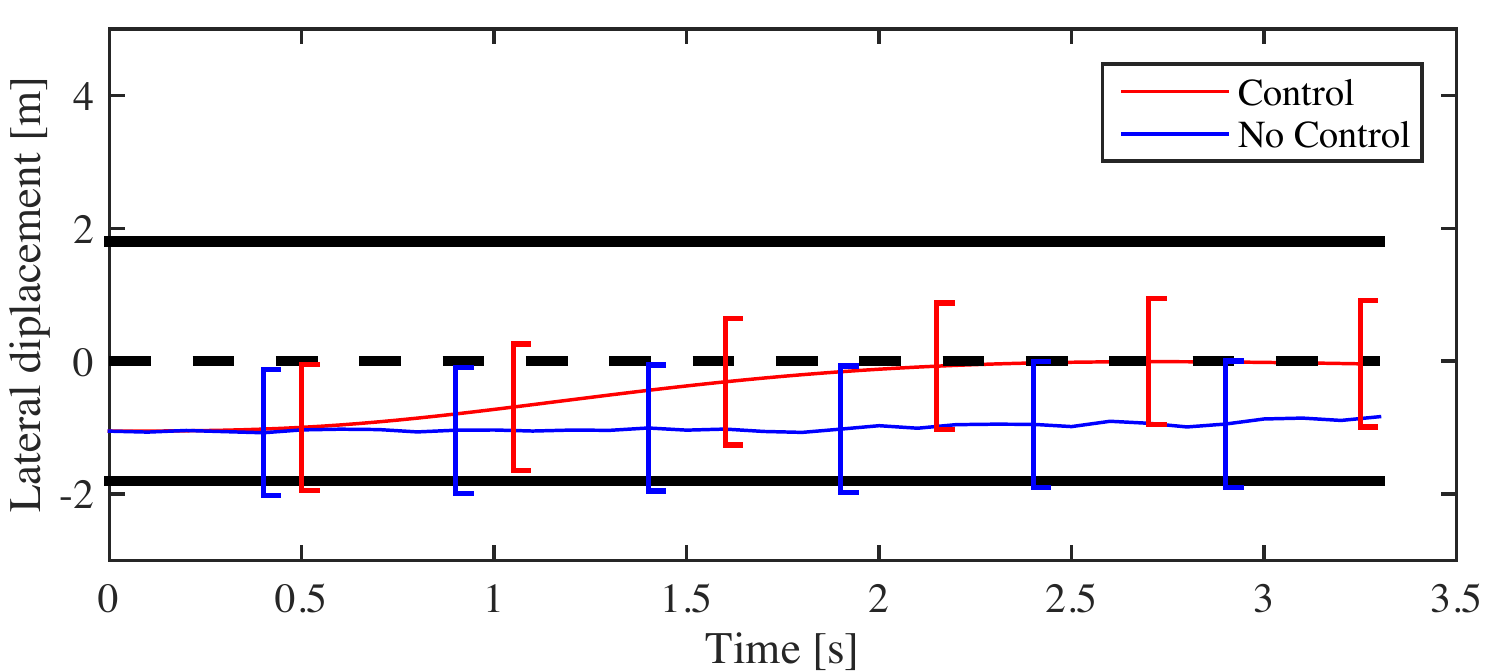}
		\caption{Right departure cases}
	\end{subfigure}
	\begin{subfigure}[t]{0.5 \textwidth}
		\centering
		\includegraphics[scale = 0.56]{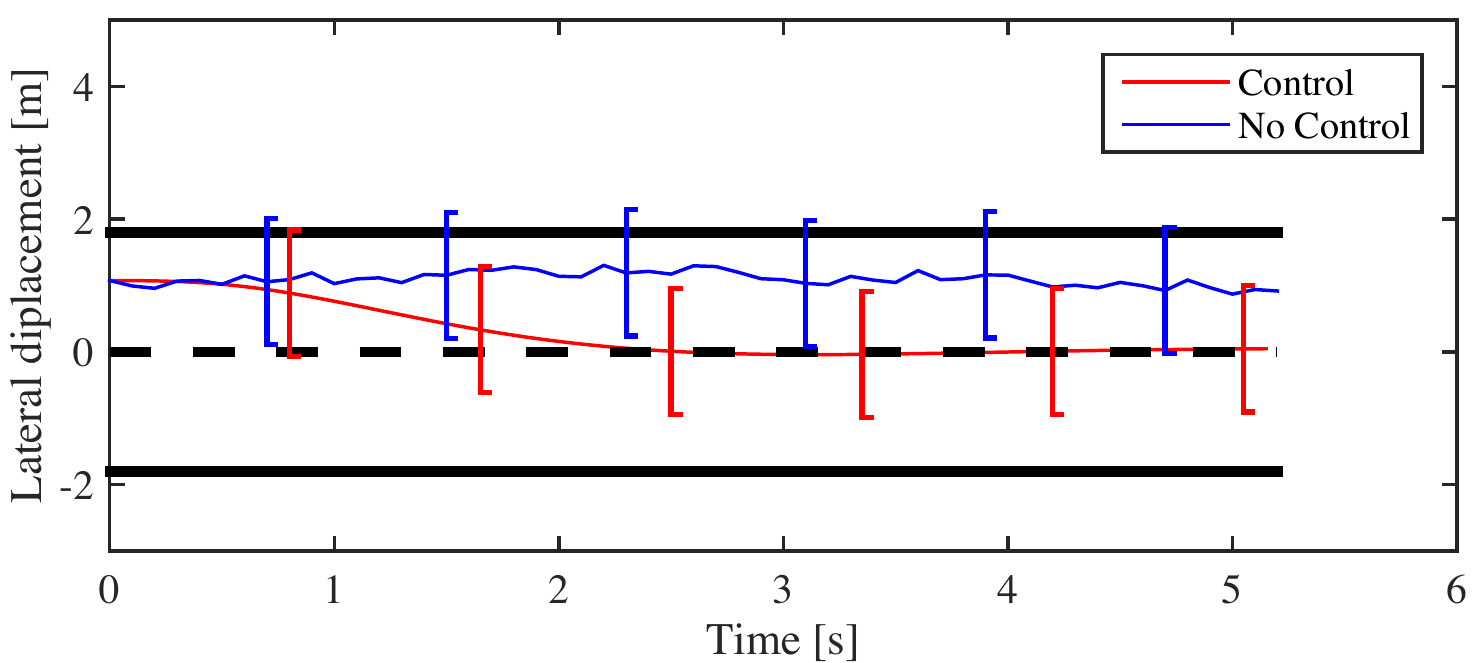}
		\caption{Left departure cases}
	\end{subfigure}
	\caption{Examples of the comparison results of the lane departure events with and without controller.}
	\label{Result_Ex}
\end{figure}

\begin{figure}[h!]
	\centering
	\includegraphics[width = 0.48\textwidth]{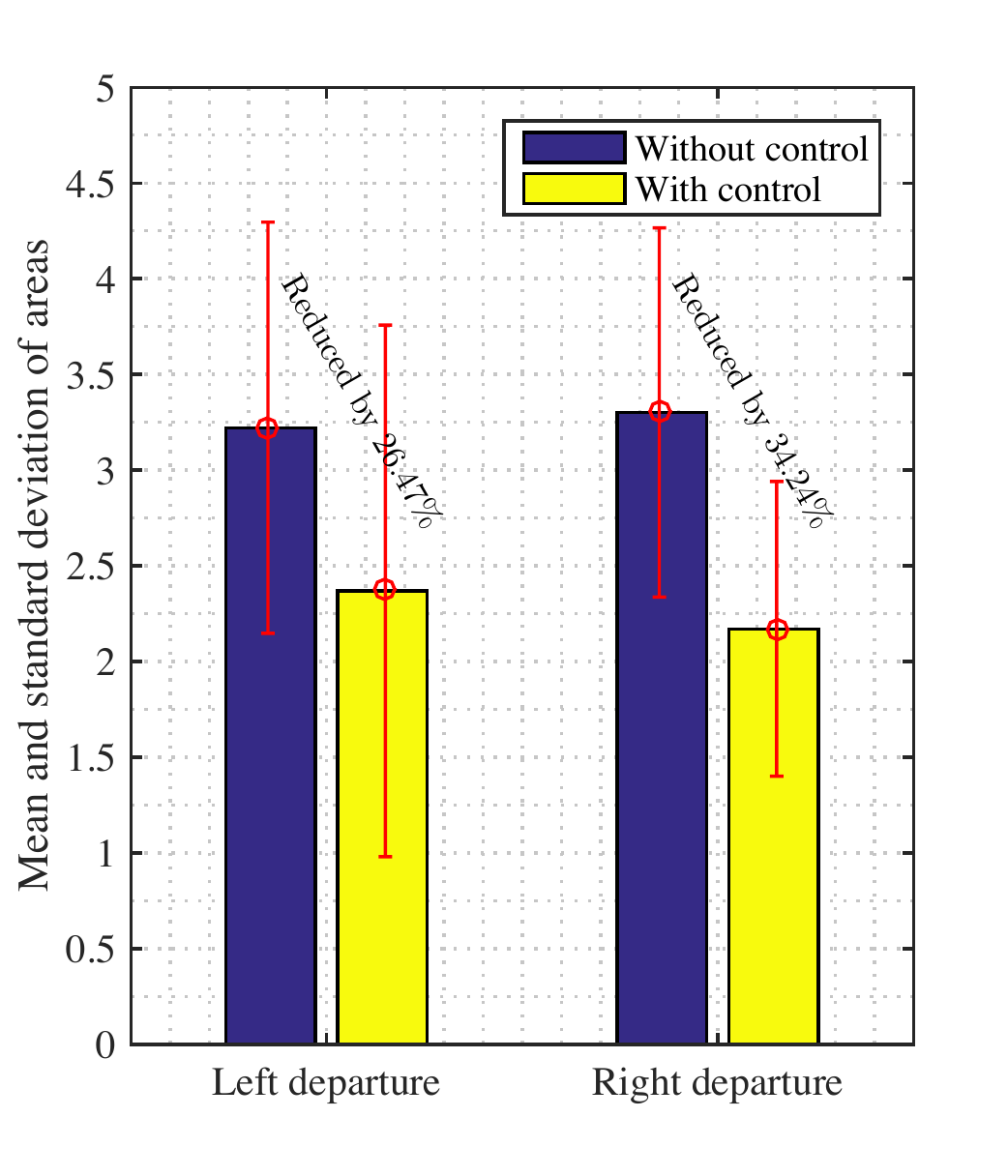}
	\caption{The mean and standard deviations of departure covering areas for 200 left and right departure events, respectively.}
	\label{Result_area}
\end{figure}

Fig. \ref{Illus_ex} illustrates the lane departure event (green solid line) generated from the stochastic model and the control results (green dash line). The red point represents the trigger point where the control condition is valid. 

The simulation results are shown in Fig. \ref{Result_Ex}. The solid black line is the lane boundary; the dashed black line is the center of the driving lane. The blue line is the lane departure trajectory generated from the stochastic model; the red line is the trajectory of the vehicle with the designed controller. In Fig. \ref{Result_Ex}, we draw the trajectory only after the controller is triggered, i.e., the trajectory after the red point in Fig. \ref{Illus_ex}. It is shown that the learning-based stochastic model regenerated the lane departure events that can be used in the controller evaluation. 

To clearly show the evaluation performance of our proposed approach, the area between the vehicle trajectory and road centerline for each departure event is computed by

\begin{equation}
S = \int_{t_{start}}^{t_{end}}   |e_{y}(t)|  dt
\end{equation}
where $ t_{start} $ and $ t_{end} $ are the start and end time point of controller, respectively. Thus, a smaller value of $ S $ indicates that the vehicle is tracking the lane center better.

\begin{figure*}[t]
	\begin{subfigure}[t]{0.68 \textwidth}
		\centering
		\includegraphics[scale = 0.7]{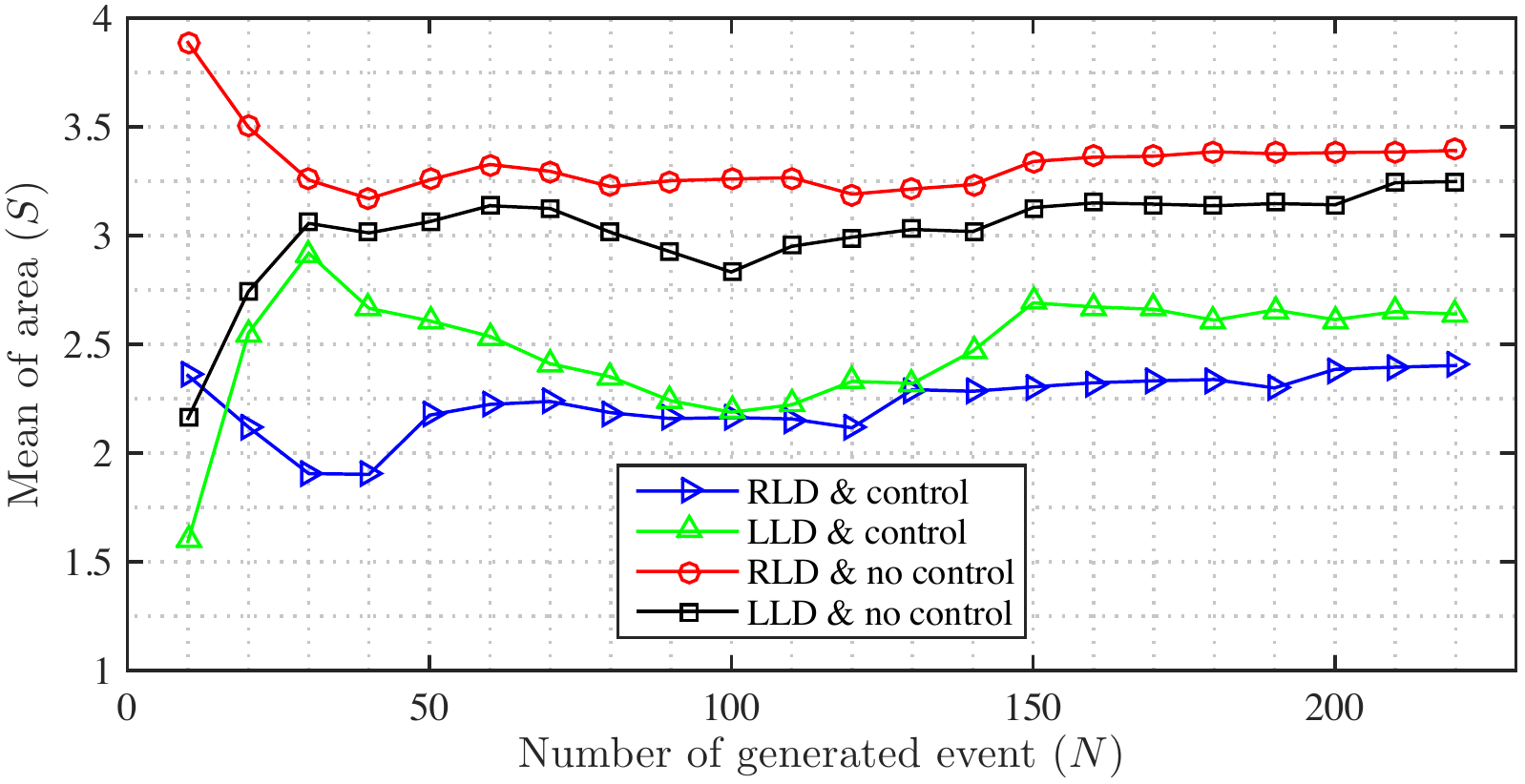}
	\end{subfigure}
	\begin{subfigure}[t]{0.28 \textwidth}
		\centering
		\includegraphics[scale = 0.7]{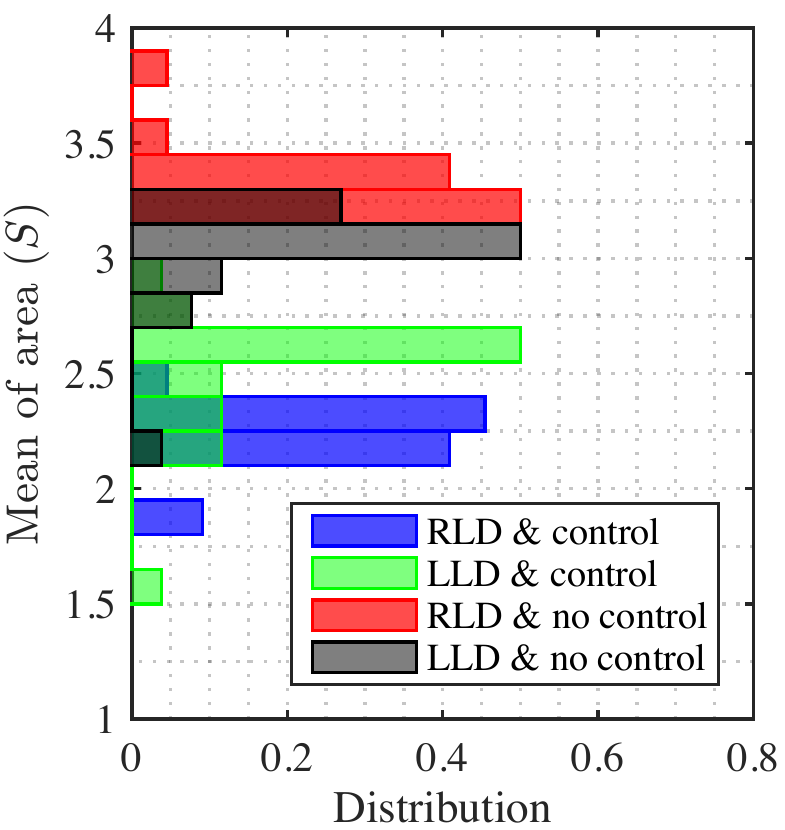}
	\end{subfigure}
	\caption{The statistical results for different number of lane departure event generated from the trained model.}
	\label{Result_stat}
\end{figure*}

Fig. \ref{Result_area} provides the statistical simulation results for 200 left lane departure events and 200 right lane departure events that were stochastically generated from the trained model. The bar represents the mean value and the red vertical line represents the standard deviation of $ S $. We note that our proposed method can generate a wide range of different kinds of lane departure events and that the controller can pull the vehicle back to the lane center when drivers depart from the lane center. The vehicle with an LDC system can reduce the departure area $ S $ by $ 26.47 \%$ and $ 34.24 \%$ for left and right departure events, respectively. Fig. \ref{Result_stat} shows the statistical results of generated lane departure events, consisting of right lane departures (RLDs) and left lane departures (LLDs), thus supporting the benefit of our proposed approach to evaluating semi-autonomous LDC systems using naturalistic driving data.

\section{Conclusions}

In this paper, we propose a framework for evaluating a semi-autonomous LDC system using a learning-based method. The proposed model can regenerate LDB to evaluate the controller for the semi-autonomous LDC system. In the stochastic model, we apply naturalistic driving data to learn the model parameters based on a bounded Gaussian mixture model. To reduce computation costs, we propose a dimension reduction method by using a polynomial function with stochastic terms to characterize the LDB. Finally, a controller is designed to validate the semi-autonomous LDC system. The simulation results indicate that the proposed stochastic model is effective in evaluating the semi-autonomous LDC system.


\section*{ACKNOWLEDGMENT}

This work was funded in part by the Toyota Class Action Settlement Safety Research and Education Program. The conclusions being expressed are the authors’ only, and have not been sponsored, approved, or endorsed by Toyota or plaintiffs’ class counsel. The authors also wish to thank Scott Bogard for his help with data processing and Carol Flannagan for our many constructive discussions.

\bibliographystyle{IEEEtran}
\bibliography{Mendeley_Lane_Departure}

\begin{thebibliography}{10}
\providecommand{\url}[1]{#1}
\csname url@samestyle\endcsname
\providecommand{\newblock}{\relax}
\providecommand{\bibinfo}[2]{#2}
\providecommand{\BIBentrySTDinterwordspacing}{\spaceskip=0pt\relax}
\providecommand{\BIBentryALTinterwordstretchfactor}{4}
\providecommand{\BIBentryALTinterwordspacing}{\spaceskip=\fontdimen2\font plus
\BIBentryALTinterwordstretchfactor\fontdimen3\font minus
  \fontdimen4\font\relax}
\providecommand{\BIBforeignlanguage}[2]{{%
\expandafter\ifx\csname l@#1\endcsname\relax
\typeout{** WARNING: IEEEtran.bst: No hyphenation pattern has been}%
\typeout{** loaded for the language `#1'. Using the pattern for}%
\typeout{** the default language instead.}%
\else
\language=\csname l@#1\endcsname
\fi
#2}}
\providecommand{\BIBdecl}{\relax}
\BIBdecl

\bibitem{enache2009driver}
N.~M. Enache, M.~Netto, S.~Mammar, and B.~Lusetti, ``Driver steering assistance
  for lane departure avoidance,'' \emph{Control engineering practice}, vol.~17,
  no.~6, pp. 642--651, 2009.

\bibitem{reagan2016observed}
I.~J. Reagan and A.~T. McCartt, ``Observed activation status of lane departure
  warning and forward collision warning of honda vehicles at dealership service
  centers,'' \emph{Traffic injury prevention}, vol.~17, no.~8, pp. 827--832,
  2016.

\bibitem{Zhao2016AcceleratedTechniques}
D.~Zhao, H.~Lam, H.~Peng, S.~Bao, D.~J. LeBlanc, K.~Nobukawa, and C.~S. Pan,
  ``{Accelerated Evaluation of Automated Vehicles Safety in Lane Change
  Scenarios Based on Importance Sampling Techniques},'' \emph{IEEE Transactions
  on Intelligent Transportation Systems}, 2016.

\bibitem{Zhao2015AcceleratedData}
D.~Zhao, H.~Peng, S.~Bao, K.~Nobukawa, D.~J. LeBlanc, and C.~S. Pan,
  ``{Accelerated evaluation of automated vehicles using extracted naturalistic
  driving data},'' in \emph{Proceeding for 24th International Symposium of
  Vehicles on Road and Tracks}, 2015.

\bibitem{Zhao2016AcceleratedManeuvers}
\BIBentryALTinterwordspacing
D.~Zhao, X.~Huang, H.~Peng, H.~Lam, and D.~J. Leblanc, ``{Accelerated
  Evaluation of Automated Vehicles in Car-Following Maneuvers},'' \emph{ArXiv},
  2016. [Online]. Available: \url{http://arxiv.org/abs/1607.02687}
\BIBentrySTDinterwordspacing

\bibitem{harding2014vehicle}
J.~Harding, G.~Powell, R.~Yoon, J.~Fikentscher, C.~Doyle, D.~Sade, M.~Lukuc,
  J.~Simons, and J.~Wang, ``{Vehicle-to-vehicle communications: Readiness of
  V2V technology for application},'' Tech. Rep., 2014.

\bibitem{wanghuman}
W.~Wang, J.~Xi, C.~Liu, and X.~Li, ``Human-centered feed-forward control of a
  vehicle steering system based on a driver's path-following characteristics,''
  \emph{IEEE Transactions on Intelligent Transportation Systems}, DOI:
  10.1109/TITS.2016.2606347.

\bibitem{Huang2016UsingScenario}
\BIBentryALTinterwordspacing
Z.~Huang, D.~Zhao, H.~Lam, and D.~J. LeBlanc, ``{Accelerated Evaluation of
  Automated Vehicles Using Piecewise Mixture Models},'' \emph{submitted to IEEE
  Transactions on Intelligent Transportation Systems}, 7 2017. [Online].
  Available: \url{http://arxiv.org/abs/1701.08915}
\BIBentrySTDinterwordspacing

\bibitem{bezzina2014safety}
D.~Bezzina and J.~Sayer, ``{Safety pilot model deployment: Test conductor team
  report},'' \emph{Report No. DOT HS}, vol. 812, p. 171, 2014.

\bibitem{stein2003vision}
G.~P. Stein, O.~Mano, and A.~Shashua, ``{Vision-based ACC with a single camera:
  bounds on range and range rate accuracy},'' in \emph{Intelligent vehicles
  symposium, 2003. Proceedings. IEEE}.\hskip 1em plus 0.5em minus 0.4em\relax
  IEEE, 2003, pp. 120--125.

\bibitem{jonas2003bounded}
J.~Lindblom and J.~Samuelsson, ``{Bounded Support Gaussian Mixture Modeling of
  Speech Spectra},'' \emph{IEEE Transactions on Speech and Audio Processing},
  vol.~11, no.~1, pp. 88--99, 2003.

\bibitem{nguyen2014bounded}
T.~M. Nguyen, Q.~M.~J. Wu, and H.~Zhang, ``{Bounded generalized Gaussian
  mixture model},'' \emph{Pattern Recognition}, vol.~47, no.~9, pp. 3132--3142,
  2014.

\bibitem{Lee2012EMData}
G.~Lee and C.~Scott, ``{EM algorithms for multivariate Gaussian mixture models
  with truncated and censored data},'' \emph{Computational Statistics and Data
  Analysis}, vol.~56, no.~9, pp. 2816--2829, 2012.

\bibitem{Box2015TimeControl}
G.~Box, G.~Jenkins, G.~Reinsel, and G.~Ljung, \emph{{Time series analysis:
  forecasting and control}}, 2015.

\end{thebibliography}

\end{document}